\documentclass[preprint,showpacs,showkeys,endfloats]{revtex4}%
\usepackage{amsfonts}
\usepackage{float}
\usepackage{amsmath}
\usepackage{amssymb}
\usepackage{graphicx}
\begin{document}
\preprint{Preprint }
\title[The extended XY model with long-range interactions]{Quantum phase transitions of the extended \ isotropic XY model with long-range
interactions }
\author{F. G. Ribeiro}
\affiliation{Departamento de F\'{\i}sica, Universidade Federal do Piau\'{\i}, Campus
Ministro Petr\^{o}nio Portela,64049-550, Teresina, Piau\'{\i}, Brazil.}
\author{J. P. de Lima}
\affiliation{Departamento de F\'{\i}sica, Universidade Federal do Piau\'{\i}, Campus
Ministro Petr\^{o}nio Portela,64049-550, Teresina, Piau\'{\i}, Brazil.}
\author{L. L. Gon\c{c}alves}
\affiliation{Departamento de Engenharia Metal\'{u}rgica e de Materiais, Universidade
Federal do Cear\'{a}, Campus do Pici, Bloco 714, 60455-760, Fortaleza,
Cear\'{a}, Brazil}
\email{lindberg@fisica.ufc.br}
\author{}

\begin{abstract}
The one-dimensional extended isotropic XY model (s=1/2) in a transverse field
with uniform long-range interactions among the \textit{z} components of the
spin is considered. The model is exactly solved by introducing the gaussian
and Jordan-Wigner transformations, which map it in a non-interacting fermion
system. The partition function can be determined in closed form at arbitrary
temperature and for arbitrary multiplicity of the multiple spin interaction.
From this result all relevant thermodynamic functions are obtained and, due to
the long-range interactions, the model can present classical and quantum
transitions of first- and second-order. The study of its critical behavior is
restricted for the quantum transitions, which are induced by the transverse
field at $T=0.$ The phase diagram is explicitly obtained for multiplicities
$p=2,3,4$ and $\infty,$ as a function of the interaction parameters, and, in
these cases, the critical behavior of the model is studied\textbf{ }in detail.
Explicit results are also presented for the induced magnetization and
isothermal susceptibility $\chi_{T}^{zz}$, and a detailed analysis is also
carried out for the static longitudinal $\langle S_{j}^{z}S_{l}^{z}\rangle$
and transversal $\langle S_{j}^{x}S_{l}^{x}\rangle$ correlation functions. The
different phases presented by the model can be characterized by the spatial
decay of the these correlations, and from these results some of these can be
classified as quantum spin liquid phases. The static critical exponents and
the dynamic one, $z,$ have also been determined, and it is shown that, besides
inducing first order phase transition, the long-range interaction also changes
the universality class the model.

\end{abstract}
\date{08/12/2009}

\eid{identifier}
\maketitle

\section{Introduction}

The study of the quantum phase transitions(QPTs) has been the object of many
theoretical and experimental investigations\cite{Schadev 2000}, since they
play an essential role in understanding the low temperature properties of the
materials\cite{Coleman 2005}. In particular, for low dimension magnetic
materials, the study of quantum spin chains has provided much insight in this
direction\cite{Schadev 2008}. Motivated by this, in the last years, others
tools have been used to investigate QPTs in spin chains, such as, quantum
discord\cite{Raoul 2008}, geometric phases\cite{Carollo 2005, Zhu 2006} and
quantum fidelity\cite{Zanardi 2006}, where the last two concepts were unified
in the approach of the geometric tensors\cite{Venuti 2007}. All these tools
have only one purpose, that is, to point out the existence of the quantum
critical behavior. On the other hand, the QPTs are also directly related to
quantum entanglement, which has been object of  study in  quantum
computation\cite{Amico 2006}. Therefore, the study of spin systems is of great
importance for understanding the behavior of the materials in low temperature regime.

Among the quantum spin models, the one-dimensional XY model introduced by Lieb
\textit{et al}.\cite{Lieb 1961}, with its generalized classes, have attracted
much interest in the last decades. In particular, the models with multiple
spin (see Refs.\cite{Titvinidze 2003, Krokhmalskii2008}, and references
therein) and long-range interactions(see Refs. \cite{Lenilson 2005, Bouchett
2008}, and references therein) have been object of intensive investigations.
As pointed out by Derzhko\cite{Derzhko 2009} and in the references therein,
quantum spin systems with multiple spin interactions work as effective spin
models for the standard Hubbard model under certain conditions. On the other
hand, models with long-range interactions are important to understand the
process of quantum information in spin chains as well as for the study of the
classical and/or quantum crossover, as it was explained by de Lima and
Gon\c{c}alves.\cite{de Lima 2008}. Another importance of the long-range
interaction is that it can induce first-order QPTs, which play an important
role in the quantum critical behaviour\cite{Pfleiderer 2005}.

Therefore, in this paper, we will study the effect of the long-range
interaction on the quantum critical behaviour of the extended one-dimensional
XY-model with multiple spin interactions\cite{Suzuki 1971}. The solution of
the model can be obtained exactly, at arbitrary temperatures in a
one-dimensional and the main purpose of the paper is to study the quantum
critical behaviour of the model.

We analyze these two kind of interactions, where we will show the role of each
one, that is, we will show that the long-range interactions induce first and
second order QPTs and that their presence change the universality class of the
model, as we will also verify the scaling relations proposed by Continentino
and Ferreira\cite{Continentino 2004} for first-order QPTs. While for the
multisite interaction among spins, we will show the presence of many different
kinds of quantum spin liquid phases, which after some time returned to be a
relevant topic in the present research\cite{Lee 2008}.

In Sec. II we introduce the model and obtain its exact solution by means of
\ Jordan-Wigner fermionization and the integral \ Gaussian transformation, and
present its basic results, such as, the spectrum of the energy, magnetization
at arbitrary temperatures. The quantum phase diagrams, as function of the
interaction parameters, are presented and discussed in Sec. III, and in Sec.
IV we study the scaling behaviour of the longitudinal and transversal
correlations functions on the different phases presented by the model. In Sec.
V we evaluate the critical exponents, and discuss the change in the
universality class due the presence of the long-range interactions. Finally,
in Sec. VI we summarize the main results.

\section{THE MODEL AND BASIC RESULTS}

We consider the one-dimensional isotropic extended XY model ($s=1/2$) with
uniform long-range interactions among the $z$ components of the spins, \ whose
Hamiltonian is explicitly given by {
\begin{align}
\mathcal{H}  &  =-J_{1}\sum_{j=1}^{N}(S_{j}^{x}S_{j+1}^{x}+S_{j}^{y}%
S_{j+1}^{y})-\frac{I}{N}\sum_{j=1}^{N}\sum_{l=1}^{N}S_{j}^{z}S_{l}^{z}%
-h\sum_{j=1}^{N}S_{j}^{z}-\nonumber\\
&  -J_{2}\sum_{j=1}^{N}\sum_{\kappa=2}^{p}(S_{j}^{x}S_{j+\kappa}^{x}+S_{j}%
^{y}S_{j+\kappa}^{y})S_{j+1}^{z}S_{j+2}^{z}...S_{j+\kappa-1}^{z},
\label{hamiltonian}%
\end{align}
{where the parameters $J_{1}$, $J_{2}$ are the exchange coupling between
nearest neighbors, }}$I${{ the uniform long-range interaction among the }}%
$z${{ components, }}$p${{ {is the multiplicity of the multiple spin
interaction, }and }}$N${{ is the number of sites of the lattice.}}

{{Introducting the Jordan-Wigner transformation
\begin{equation}
S_{j}^{+}=\left[  \text{exp}\left(  i\pi\sum_{l=1}^{j-1}c_{l}^{\dagger}%
c_{l}\right)  \right]  c_{j}^{\dagger}\hspace{0.1cm};\hspace{0.5cm}S_{j}%
^{-}=c_{j}\left[  \text{exp}\left(  -i\pi\sum_{l=1}^{j-1}c_{l}^{\dagger}%
c_{l}\right)  \right]  , \label{jordanwigner}%
\end{equation}%
\begin{equation}
S_{j}^{z}=c_{j}^{\dagger}c_{j}-\frac{1}{2}, \label{sz}%
\end{equation}
where $c_{j}$ and $c_{j}^{\dagger}$ are fermion operators, the Hamiltonian can
be written in the well known decomposed form(\cite{siskens 1974} also
references therein)}}%
\begin{equation}
\mathcal{H}=\mathcal{H}^{+}P^{+}+\mathcal{H}^{-}P^{-}, \label{decomposed}%
\end{equation}
where%
\begin{align}
\mathcal{H}^{\pm}  &  =-\frac{J_{1}}{2}\sum_{j=1}^{N-1}\left(  c_{j}^{\dagger
}c_{j+1}+c_{j+1}^{\dagger}c_{j}\right)  -\left(  h-I\right)  \sum_{j=1}%
^{N}c_{j}^{\dagger}c_{j}-\frac{I}{N}\sum_{j=1}^{N}\sum_{l=1}^{N}c_{j}%
^{\dagger}c_{j}c_{l}^{\dagger}c_{l}\label{H+-}\\
&  -{}\frac{J_{2}}{2}\sum_{j=1}^{N-p}\sum_{k=2}^{p}\left(  \frac{-1}%
{2}\right)  ^{k-1}\left(  c_{j}^{\dagger}c_{j+k}+c_{j+k}^{\dagger}%
c_{j}\right)  +\frac{N}{2}\left(  h-\frac{I}{2}\right) \nonumber\\
&  \pm\frac{J_{1}}{2}\left(  c_{N}^{\dagger}c_{1}+c_{1}^{\dagger}c_{N}\right)
\pm{}\frac{J_{2}}{2}\sum_{l=1}^{p}\sum_{\substack{k=2\\p-l\geq2}}^{p-l}\left(
\frac{-1}{2}\right)  ^{k-1}\left(  c_{N-p+l}^{\dagger}c_{l-p+k}+c_{l-p+k}%
^{\dagger}c_{N-p+l}\right)  \pm\nonumber\\
&  \pm\frac{J_{2}}{2}\sum_{l=1}^{p}\sum_{\substack{k=p-l+1\\k\geq2}%
}^{p}\left(  \frac{-1}{2}\right)  ^{k-1}\left(  c_{N-p+l}^{\dagger}%
c_{l-p+k}+c_{l-p+k}^{\dagger}c_{N-p+l}\right)  ,\nonumber
\end{align}
{{{ }}}and%
\begin{equation}
P^{\pm}=\frac{I\pm P}{2}, \label{P+-}%
\end{equation}
with $P$ given by%
\begin{equation}
P=\exp\left(  i\pi\sum_{l=1}^{N}c_{l}^{\dagger}c_{l}\right)  .
\label{Pexpression}%
\end{equation}

As it is also well known, the Hamiltonian $\mathcal{H}^{\pm}$ can be
diagonalized by imposing \ antiperiodic (for $\mathcal{H}^{+})$ and periodic
(for $\mathcal{H}^{-})$ boundary conditions, and, in the thermodynamic limit,
the static properties can be described by $\mathcal{H}^{-}$ \cite{siskens
1974,capel 1977,goncalves 1977}. Therefore, since we are interested in the
determination of the static properties, we will identify the Hamiltonian of
the system with $\mathcal{H}^{-}.$ By taking into account that {the long-range
interaction term commutes with the Hamiltonian, }the partition can be written
in the form

{%
\begin{align}
\mathcal{Z_{N}}  &  =\text{exp}\left[  -\frac{\beta N}{2}\left(  h-\frac{I}%
{2}\right)  \right]  Tr\left\{  \text{exp}\left[  \frac{\beta J_{1}}{2}%
\sum_{j=1}^{N}\left(  c_{j}^{\dagger}c_{j+1}+c_{j+1}^{\dagger}c_{j}\right)
+\right.  \right.  {}\nonumber\\
&  {}\left.  \left.  +\beta\left(  h-I\right)  \sum_{j=1}^{N}c_{j}^{\dagger
}c_{j}+\frac{\beta J_{2}}{2}\sum_{j=1}^{N}\sum_{\kappa=2}^{p}\left(  \frac
{-1}{2}\right)  ^{\kappa-1}\left(  c_{j}^{\dagger}c_{j+\kappa}+c_{j+\kappa
}^{\dagger}c_{j}\right)  \right]  \right.  {}\nonumber\\
&  {}\left.  \times\text{exp}\left[  \frac{\beta I}{N}\sum_{j=1}^{N}\sum
_{l=1}^{N}c_{j}^{\dagger}c_{j}c_{l}^{\dagger}c_{l}\right]  \right\}  ,
\label{partitionfunction}%
\end{align}
{where $\beta=1/k_{B}T$ and $T$ is the temperature.}}\textbf{ }{{Introducting
the Gaussian transformation\cite{amit1984}
\begin{equation}
\text{exp}(a^{2})=\frac{1}{\sqrt{2\pi}}\int_{-\infty}^{\infty}\text{exp}%
\left(  -\frac{x^{2}}{2}+\sqrt{2}ax\right)  dx, \label{gaussiantransformation}%
\end{equation}
the partition function can be written in the integral representation as}%
\begin{align}
\mathcal{Z_{N}}  &  =\text{exp}\left[  -\frac{N}{2}\left(  \overline{h}%
-\frac{\overline{I}}{2}\right)  \right]  \sqrt{\frac{N}{2\pi}}\int_{-\infty
}^{\infty}\exp\left(  -\frac{N\overline{x}^{2}}{2}\right)  Tr\left\{
\text{exp}\left[  \frac{\overline{J}_{1}}{2}\sum_{j=1}^{N}\left(
c_{j}^{\dagger}c_{j+1}+c_{j+1}^{\dagger}c_{j}\right)  +\right.  \right.
{}\nonumber\\
&  {}\left.  +\left(  \overline{h}-\overline{I}+2\sqrt{\overline{I}}%
\overline{x}\right)  \sum_{j=1}^{N}c_{j}^{\dagger}c_{j}+\frac{\overline{J}%
_{2}}{2}\sum_{j=1}^{N}\sum_{\kappa=2}^{p}\left(  \frac{-1}{2}\right)
^{\kappa-1}\left(  c_{j}^{\dagger}c_{j+\kappa}+c_{j+\kappa}^{\dagger}%
c_{j}\right)  \right]  d\overline{x}, \label{integrepresentation}%
\end{align}
}where $\overline{x}\equiv x/\sqrt{N},\overline{J}_{1}\equiv\beta
J_{1},\overline{J}_{2}\equiv\beta J_{2},\overline{h}\equiv\beta h${{ and }%
}$\overline{I}=\beta I.$

Introducing the canonical transformation%
\begin{equation}
c_{j}=\frac{1}{\sqrt{N}}\sum_{q=1}^{N}\exp\left(  ijq\right)  c_{q}%
\ \ \ \ \ \text{and}\ \ \ \ \ \ \ c_{j}^{\dagger}=\frac{1}{\sqrt{N}}\sum
_{q=1}^{N}\exp\left(  -ijq\right)  c_{q}^{\dagger}, \label{fourier}%
\end{equation}
with \ $q=2\pi n/N,$ $n=1,2,...,N,$ Eq. (\ref{integrepresentation}) \ can be
written in the form%
\begin{equation}
{\mathcal{Z_{N}}=C(\beta)\int_{-\infty}^{\infty}\exp\left(  -\frac
{N\overline{x}^{2}}{2}\right)  \zeta\left(  \overline{x}\right)  d\overline
{x},} \label{compa}%
\end{equation}
where%
\begin{equation}
C(\beta)=\sqrt{\frac{N}{2\pi}}\text{exp}\left[  -\frac{N}{2}\left(
\overline{h}-\frac{\overline{I}}{2}\right)  \right]
,\ \ \ \ \ \ \ \ \ \ \ \ \zeta\left(  \overline{x}\right)  =Tr\left\{
\exp\left[  \sum_{q}\overline{\epsilon}_{q}\left(  \overline{x}\right)
c_{q}^{\dagger}c_{q}\right]  \right\}  , \label{constants}%
\end{equation}
with%
\begin{equation}
\overline{\epsilon}_{q}\left(  \overline{x}\right)  =\beta\left[  J_{1}%
\cos(q)+J_{2}\sum_{\kappa=2}^{p}\left(  -\frac{1}{2}\right)  ^{\kappa-1}%
\cos(\kappa q)+h-I+\sqrt{\frac{2I}{\beta}}\overline{x}\right]  .
\label{energy}%
\end{equation}

In the thermodynamic limit, $N\rightarrow\infty$, we use the Laplace's
method\cite{murray 1984} to evaluate the partition function, and
${\mathcal{Z_{N}}}$ can be written in the\ form%
\begin{equation}
{\mathcal{Z_{N}=}}\frac{\exp\left[  -\frac{N}{2}\left(  \overline{h}%
-\frac{\overline{I}}{2}\right)  +Ng(\overline{x}_{0})\right]  }{\mid
g^{\prime\prime}(\overline{x}_{0})\mid^{1/2}}, \label{final}%
\end{equation}
where%
\begin{equation}
g(\overline{x}_{0})=-\frac{\overline{x}_{0}^{2}}{2}+\frac{1}{N}\sum_{q}%
\ln\left[  1+\exp\left(  \overline{\epsilon}_{q}\left(  \overline{x}%
_{0}\right)  \right)  \right]  , \label{funlaplace}%
\end{equation}
with $g(\overline{x}_{0})\ $satisfying the conditions%
\begin{equation}
g^{\prime}(\overline{x}_{0})=0,\text{ \ \ \ \ \ and\ \ \ \ \ \ \ }%
g^{\prime\prime}(\overline{x}_{0})<0, \label{condition}%
\end{equation}
and $\overline{x}_{0}$ explicitly given by%
\begin{equation}
\overline{x}_{0}=\frac{\sqrt{2\overline{I}}}{N}\sum_{q=1}^{N}\frac{1}%
{1+\exp(-\overline{\epsilon}_{q}\left(  \overline{x}_{0}\right)  )}.
\label{xo}%
\end{equation}
Therefore, from Eq.(\ref{final}) we can write the Helmholtz free energy as%
\begin{equation}
F_{N}=\frac{N}{2}\left(  h-\frac{I}{2}\right)  -Nk_{B}Tg(\overline{x}%
_{0})+\frac{k_{B}T}{2}\ln\mid g^{\prime\prime}(\overline{x}_{0})\mid.
\label{functional}%
\end{equation}

Taking into account that the magnetization per site $M^{z}$ can be written in
the form%
\[
M^{z}=\frac{1}{N}\sum_{j}\langle S_{j}^{z}\rangle=\frac{1}{N}\sum_{q}\langle
c_{q}^{\dagger}c_{q}\rangle-\frac{1}{2}=\frac{1}{N}\sum_{q}\frac{1}%
{1+\exp\left[  -\overline{\epsilon}_{q}\left(  \overline{x}\right)  \right]
}-\frac{1}{2},
\]
by using Eq.(\ref{xo}), we can express $\overline{x}_{0}$ in terms of $M^{z}$
in the form
\begin{equation}
\overline{x}_{0}=\sqrt{2\overline{I}}\left(  M^{z}+\frac{1}{2}\right)
,\label{linkxmz}%
\end{equation}
\ and from this result it follows that the functional of the Helmholtz free
energy per site is given by%
\begin{equation}
f=\frac{h}{2}+IM^{z}\left(  1+M^{z}\right)  +\frac{k_{B}T}{\pi}\int_{0}^{\pi
}\ln\left(  1+\exp\left[  \overline{\epsilon}_{q}\left(  M^{z}\right)
\right]  \right)  dq,\label{funcpersite}%
\end{equation}
with%
\begin{equation}
\overline{\epsilon}_{q}\left(  M^{z}\right)  =-\overline{J}_{1}\cos
(q)-\overline{J}_{2}\sum_{\kappa=2}^{p}\left(  -\frac{1}{2}\right)
^{\kappa-1}\cos(\kappa q)-\overline{h}-2\overline{I}M^{z}.\label{energyfinal}%
\end{equation}

From Eq. (\ref{funcpersite}),\ we can \ determine the equation of state
\ imposing \ the stability conditions
\begin{align}
\frac{\partial f}{\partial M^{z}} &  =0,\text{ \ \ \ \ \ }\label{stability}\\
\text{\ \ \ \ }\frac{\partial^{2}f}{\partial M^{z2}} &  >0,
\end{align}
which leads to the result%
\begin{equation}
M^{z}=\frac{1}{2\pi}\int_{0}^{\pi}\tanh\left[  \frac{\overline{\epsilon}%
_{q}\left(  M^{z}\right)  }{2}\right]  dq.\label{state}%
\end{equation}

\section{QUANTUM CRITICAL BEHAVIOR}

In the limit $T\rightarrow0$, the functional of the Helmholtz free energy per
site[Eq. (\ref{funcpersite})] is given by%
\begin{equation}
f=\frac{h}{2}+IM^{z}\left(  1+M^{z}\right)  -\frac{1}{\pi}%
%TCIMACRO{\dint \limits_{\epsilon_{q}\left(  M^{z}\right)  <0}}%
%BeginExpansion
{\displaystyle\int\limits_{\epsilon_{q}\left(  M^{z}\right)  <0}}
%EndExpansion
\epsilon_{q}\left(  M^{z}\right)  dq, \label{funquantum}%
\end{equation}
where
\begin{equation}
\epsilon_{q}\left(  M^{z}\right)  =J_{1}\cos(q)+J_{2}\sum_{\kappa=2}%
^{p}\left(  -\frac{1}{2}\right)  ^{\kappa-1}\cos(\kappa q)+h+2IM^{z},
\label{energyequation}%
\end{equation}
and the equation of state, given by Eq. (\ref{state}), can be written in the
form%
\begin{equation}
M^{z}=\frac{1}{2\pi}\int_{0}^{\pi}sign\left[  \epsilon_{q}\left(
M^{z}\right)  \right]  dq. \label{magquantum}%
\end{equation}

The quantum phase diagram of the model, for second order phase transitions and
arbitrary $p$, can be obtained from the previous expression by imposing the
divergence of the isothermal susceptibility, $\chi_{T}^{zz}\equiv\partial
M^{z}/\partial h\rightarrow\infty,$ and for first-order phase transitions, by
using the equation of state, Eq. (\ref{state}), and by imposing the condition
\begin{equation}
f(M_{it}^{z})=f(M_{jt}^{z}),\label{condition2}%
\end{equation}

\noindent where $M_{it}^{z}$ and $M_{jt}^{z}$ are the values of the induced
magnetization at the transition.

Following Titvinidze and Japaridze\cite{Titvinidze 2003}, by introducing the
unitary transformation
\begin{equation}
S_{j}^{x,y}\rightarrow\left(  -1\right)  ^{j}S_{j}^{x,y}\text{ \ \ \ \ ;
\ }S_{j}^{z}=S_{j}^{z},\label{unitrans}%
\end{equation}
and the time reversal one
\begin{equation}
S_{j}^{x,z}\rightarrow-S_{j}^{x,z}\text{ \ \ \ \ ; \ }S_{j}^{y}=S_{j}%
^{y},\label{timetransform}%
\end{equation}
we can show that the Hamiltonian is invariant under the transformation
$J_{1}\rightarrow-J_{1}$ and $J_{2}\rightarrow-J_{2}.$ This means that only
the signs of $I$ and $J_{2}$ are relevant  in determining the critical
behaviour of the system, and the appearance of multiple phases are result of
the competition between  these interactions which induce frustration in the
system. In particular, they can induce the so-called quantum spin liquid
phases\cite{Lee 2008}, as in the case of  the model without long-range
interaction\cite{Titvinidze 2003}. Therefore, without loss of generality, we
will consider in all results presented $J_{1}>0$ only.

From Eqs. (\ref{funquantum}-\ref{magquantum}), we can show that, for arbitrary
$p,$ the system presents\ second order quantum transitions for $I\leq0$, and
first-order quantum transitions $I>0,$ as in the previously studied XY-models
with similar long-range interactions\cite{Lenilson 2005}.

Although the Eqs. (\ref{funquantum}-\ref{magquantum}) allow us to obtain the
phase diagrams for arbitrary $p$, we will only consider the cases $p=2,3,4$
and $\infty,$ since they present the main features and, in these cases,
analytical expressions can be obtained for the critical surfaces and critical
lines associated to the second order quantum transitions.

\begin{figure}
[ptb]
\begin{center}
\includegraphics[
 % width=1\columnwidth]
%natheight=8.853900in,
%natwidth=11.458800in,
height=5.5in,
width=6.in
]%
{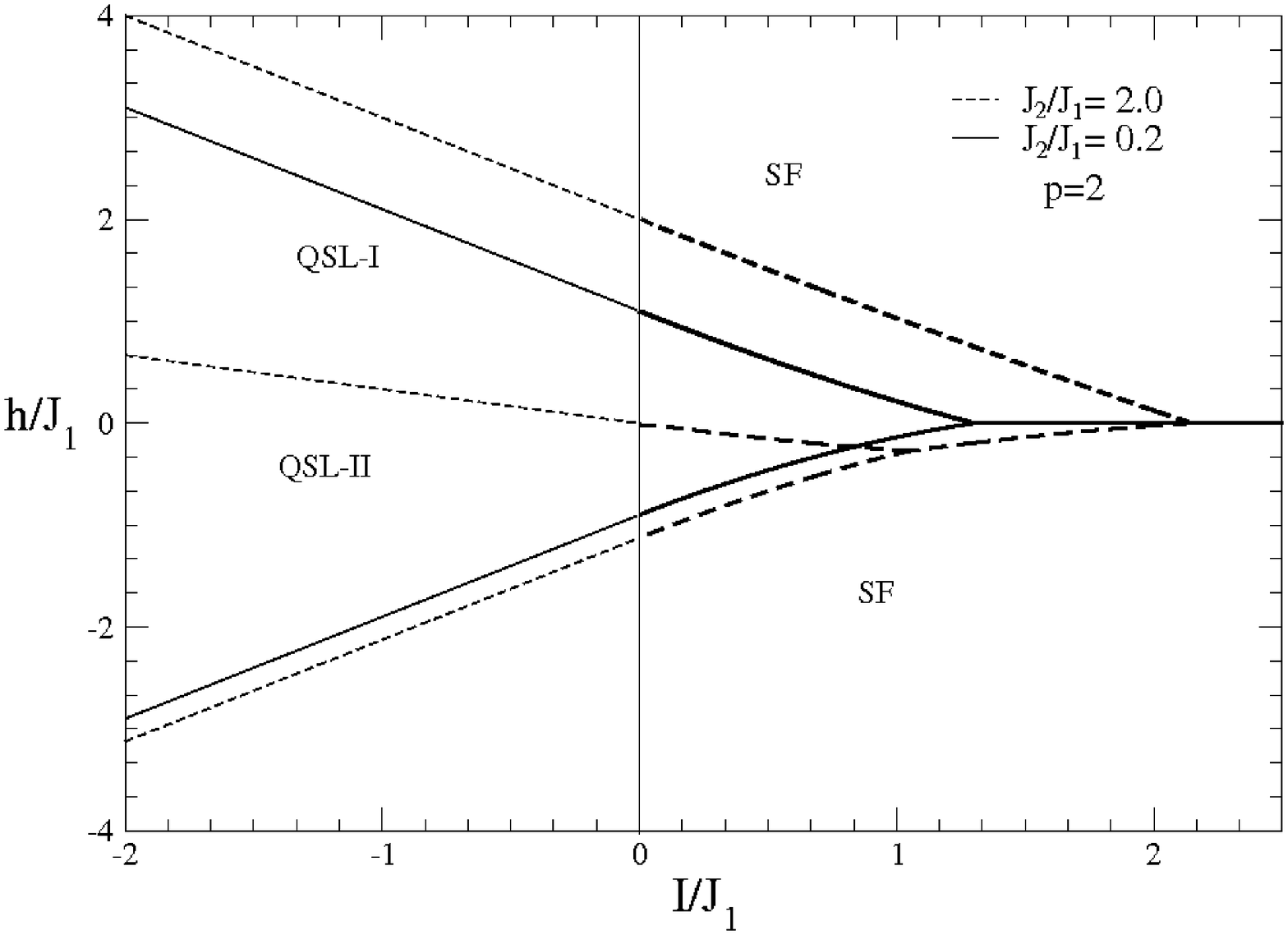}%
\caption{Phase diagram for the quantum transitions \ as a function of the
long-range interaction $I/J_{1}$ for $p=2$ and $J_{2}/J_{1}=0.2,2.0.$ For
$J_{2}/J_{1}=0.2$ there are three phases, one spin liquid phase (QSL-I) and
two saturated ferromagnetic phases(SF), and for $J_{2}/J_{1}=2.0,$ there are
four phases, two spin liquid phases (QSL-I,QSL-II) and two saturated
ferromagnetic phases(SF). The critical lines correspond to first-order phase
transitions for $I/J_{1}>0$ and to second-order phase transitions for
$I/J_{1}\leq0.$}%
\label{fig01}%
\end{center}
\end{figure}

\begin{figure}
[ptb]
\begin{center}
\includegraphics[
natheight=8.853900in,
natwidth=11.458800in,
height=4.3664in,
width=5.642in
]%
{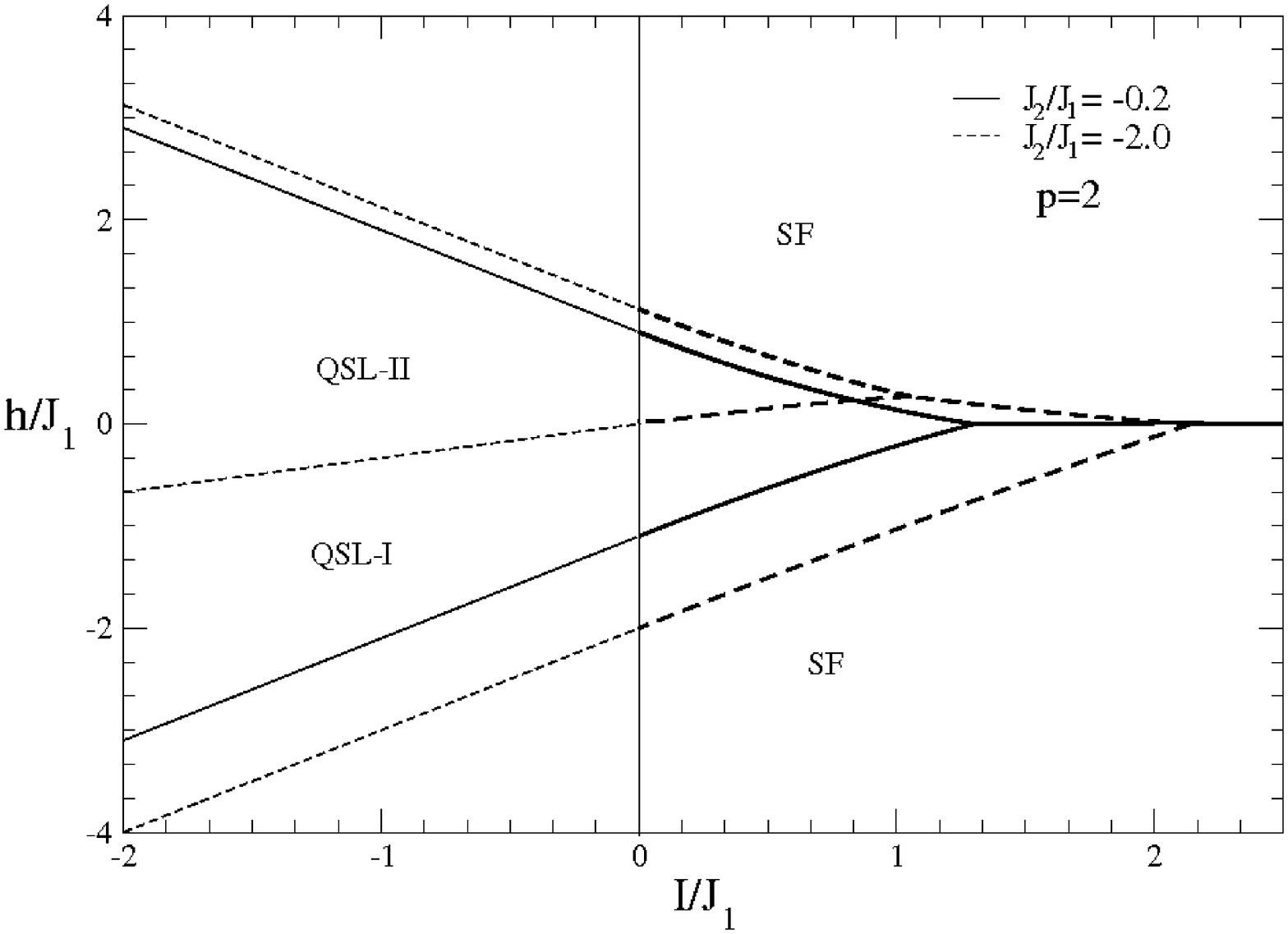}%
\caption{Phase diagram for the quantum transitions \ as a function of the
long-range interaction $I/J_{1}$ for $p=2$ and $J_{2}/J_{1}=-0.2,-2.0.$ For
$J_{2}/J_{1}=-0.2$ there are three phases, one spin liquid phase (QSL-I) and
two saturated ferromagnetic phases(SF), and for $J_{2}/J_{1}=-2.0,$ there are
four phases, two spin liquid phases (QSL-I,QSL-II) and two saturated
ferromagnetic phases(SF). The critical lines correspond to first-order phase
transitions for $I/J_{1}>0$ and to second-order phase transitions for
$I/J_{1}\leq0.$}%
\label{fig02}%
\end{center}
\end{figure}

For the case $p=2,$ which has also been studied by Titvinidze and
Japaradze\cite{Titvinidze 2003}and Krokhmalski \textit{et al}.
\cite{Krokhmalskii2008} for the model without long-range interactions, we show
in Figs. \ref{fig01} and \ref{fig02} the critical field $h/J_{1}$ as a
function of $I/J_{1}$ in the regions $J_{2}/J_{1}\geq0$ and $J_{2}/J_{1}%
\leq0,$ respectively, \ for different values $J_{2},$ which are projections of
the global phase diagram. It is worth  mentioning that in this case the
results for $J_{2}/J_{1}\leq0$ can be \ obtained from the results for
$J_{2}/J_{1}\geq0$ by introducing the transformations $h/J_{1}\rightarrow
-h/J_{1}$ and $J_{2}/J_{1}\rightarrow-J_{2}/J_{1},$ as can be verified in the
results\ shown in Figs. \ref{fig01} and \ref{fig02}.

\begin{figure}
[ptb]
\begin{center}
\includegraphics[
natheight=8.853900in,
natwidth=11.458800in,
height=4.3664in,
width=5.642in
]%
{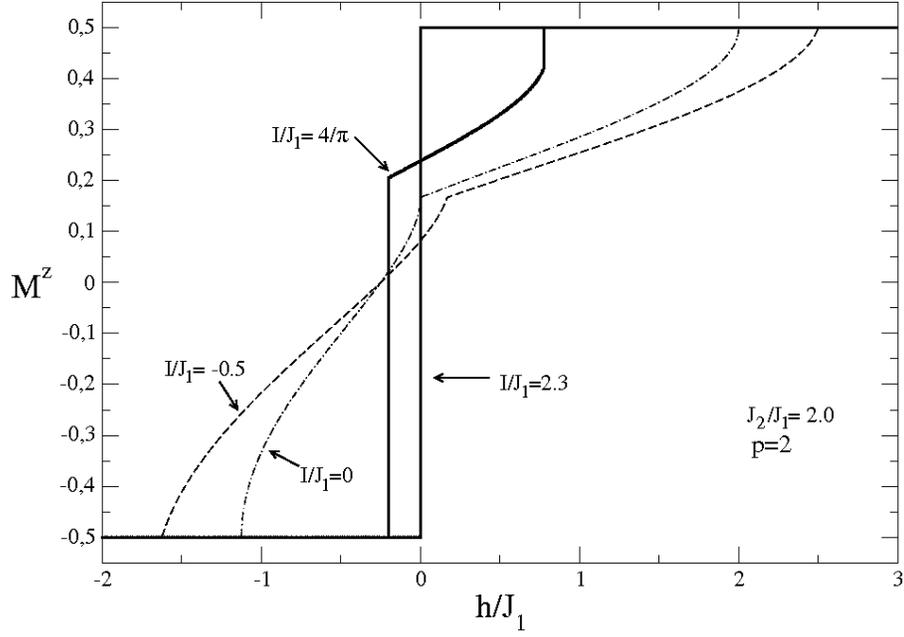}%
\caption{Magnetization $M^{z}$ as a function of $h/J_{1},$ for $p=2$,
$J_{2}/J_{1}=2.0$ and different values $I/J_{1}.$ }%
\label{fig03}%
\end{center}
\end{figure}

\begin{figure}
[ptb]
\begin{center}
\includegraphics[
natheight=8.853900in,
natwidth=11.458800in,
height=4.3664in,
width=5.642in
]%
{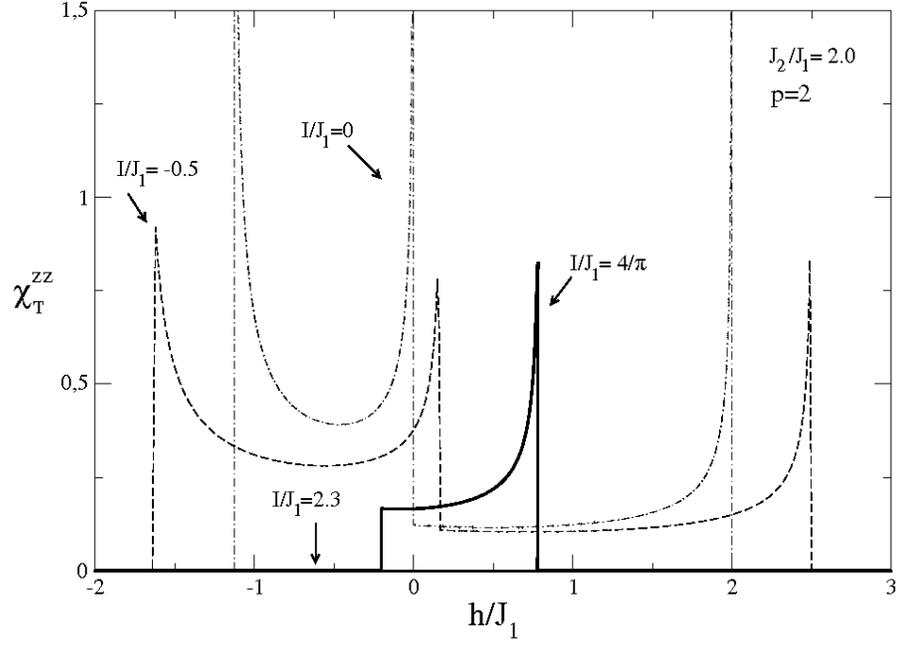}%
\caption{Isothermal susceptibility $\chi_{T}^{zz}$ as a function of $h/J_{1},$
for $p=2$, $J_{2}/J_{1}=2.0$ and different values $I/J_{1}.$}%
\label{fig04}%
\end{center}
\end{figure}

In Figs. \ref{fig03} and \ref{fig04} we present the magnetization and the
isothermal susceptibility, respectively, and from their behavior we can
conclude that for $I/J_{1}\leq0$ the system undergoes second order
transitions, and first-order transitions for $I/J_{1}>0.$

As can it be seen in Figs. \ref{fig01} and \ref{fig02}, the number of
transitions of first and second order depends on $J_{2},$ and it can also be
shown that these transitions correspond to three phases for $0\leq J_{2}%
/J_{1}\leq1/2,$and to four phases for $J_{2}/J_{1}>1/2.$ Following Titvinidze
and Japararidze\cite{Titvinidze 2003}, we can classify the intermediate
phases, which are limited by the two saturate ferromagnetic phases, as quantum
spin liquid phases\cite{Lee 2008}. As  will be shown later, these spin liquid
phases will be characterized by the spatial decay of the transversal static
correlation function $\langle S_{j}^{x}S_{l}^{x}\rangle$ and by the
modification of the oscillatory modulation of longitudinal static correlation
function $\langle S_{j}^{z}S_{l}^{z}\rangle$.

\begin{figure}
[ptb]
\begin{center}
\includegraphics[
%natheight=12.562300in,
%natwidth=16.249800in,
height=5.5in,
width=6.5in
]%
{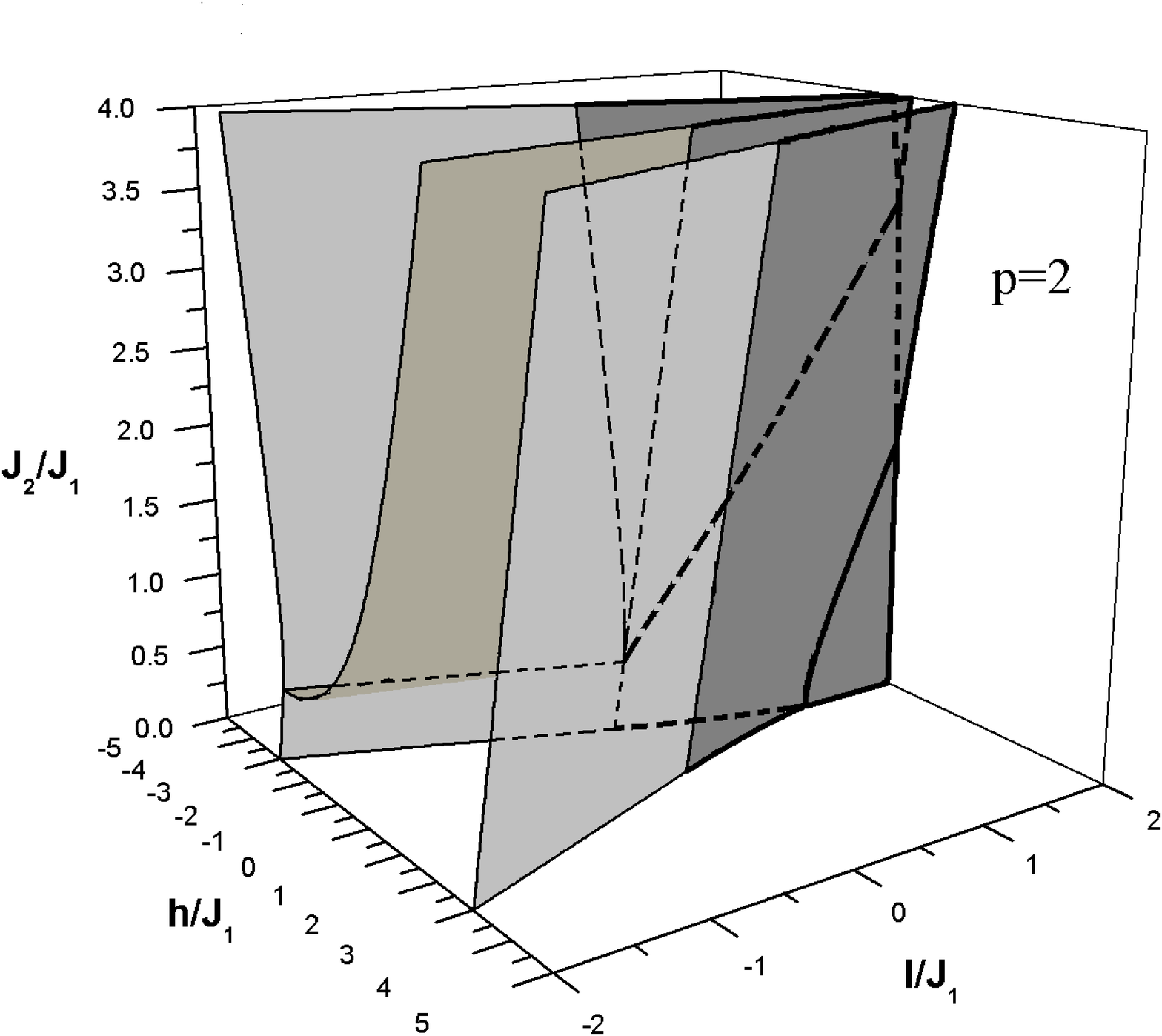}%
\caption{Global phase diagram for $p=2$ and $J_{2}/J_{1}\geq0$, as a function
of $h/J_{1}$ and $I/J_{1}$.}%
\label{fig05}%
\end{center}
\end{figure}

\begin{figure}
[ptb]
\begin{center}
\includegraphics[
%natheight=13.280900in,
%natwidth=17.187300in,
height=5.5in,
width=6.5in
]%
{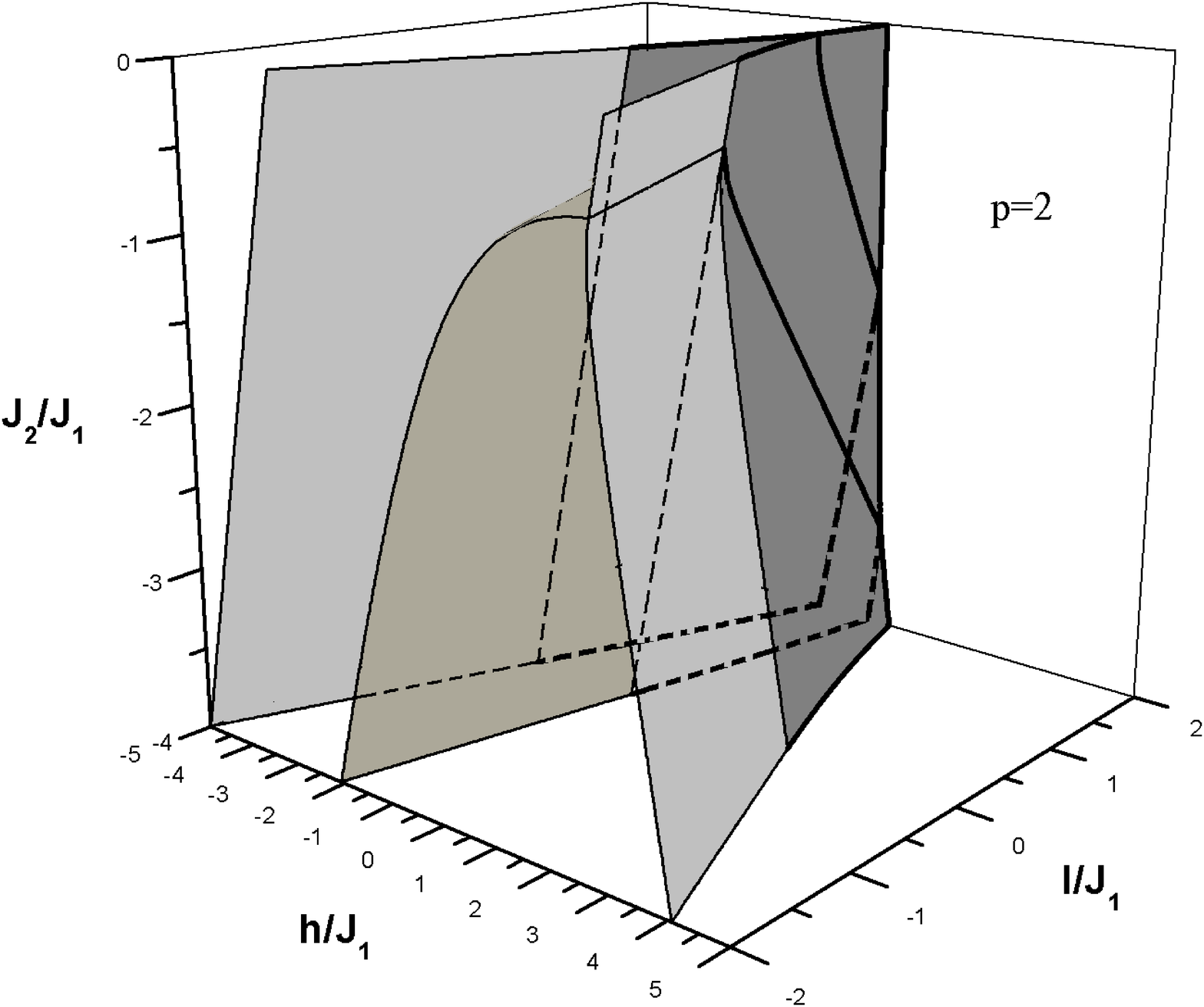}%
\caption{Global phase diagram for $p=2$ and $J_{2}/J_{1}\leq0$, as a function
of $h/J_{1}$ and $I/J_{1}$.}%
\label{fig06}%
\end{center}
\end{figure}

The global phase diagram for $p=2$ is shown in Figs. \ref{fig05} and
\ref{fig06}, for $J_{2}/J_{1}\geq0$ $\ $and $J_{2}/J_{1}\leq0,$ respectively.
For $I\leq0,$ the critical surfaces can be obtained explicitly and, for
$J_{2}/J_{1}\geq0,$ where there are four critical surfaces, which are given by
the equations%

\begin{equation}
\frac{J_{2}}{2J_{1}}-\frac{I}{J_{1}}-\frac{h}{J_{1}}+1=0,\text{ for }%
\frac{J_{2}}{J_{1}}\geq0, \label{sup21}%
\end{equation}%
\begin{equation}
\frac{J_{2}}{2J_{1}}+\frac{I}{J_{1}}-\frac{h}{J_{1}}-1=0,\text{for }0\leq
\frac{J_{2}}{J_{1}}\leq\frac{1}{2}, \label{sup22}%
\end{equation}%
\begin{equation}
-\frac{J_{2}}{2J_{1}}-\frac{J_{1}}{4J_{2}}+\frac{I}{J_{1}}-\frac{h}{J_{1}%
}=0,\text{ for }\frac{J_{2}}{J_{1}}\geq\frac{1}{2}, \label{sup23}%
\end{equation}%
\begin{equation}
\frac{J_{2}}{2J_{1}}-\frac{2IM_{1}^{z}}{J_{1}}-\frac{h}{J_{1}}+1=0,\text{ for
}\frac{J_{2}}{J_{1}}\geq\frac{1}{2}, \label{sup24}%
\end{equation}
where%
\begin{equation}
M_{1}^{z}=\frac{\arccos(\frac{J_{1}}{J_{2}}-1)}{\pi}-\frac{1}{2}, \label{mzp2}%
\end{equation}
and for $J_{2}/J_{1}\leq0$ there are four critical surfaces given by%
\begin{equation}
-\frac{J_{2}}{2J_{1}}+\frac{I}{J_{1}}+\frac{h}{J_{1}}-1=0,\text{ for }%
\frac{J_{2}}{J_{1}}\leq0, \label{supn21}%
\end{equation}%
\begin{equation}
-\frac{J_{2}}{2J_{1}}-\frac{I}{J_{1}}+\frac{h}{J_{1}}+1=0,\text{ for }%
-\frac{1}{2}\leq\frac{J_{2}}{J_{1}}\leq0, \label{supn22}%
\end{equation}%
\begin{equation}
\frac{J_{2}}{2J_{1}}+\frac{J_{1}}{4J_{2}}-\frac{I}{J_{1}}+\frac{h}{J_{1}%
}=0,\text{ for }\frac{J_{2}}{J_{1}}\leq-\frac{1}{2}, \label{supn23}%
\end{equation}%
\begin{equation}
-\frac{J_{2}}{2J_{1}}-\frac{2IM_{2}^{z}}{J_{1}}+\frac{h}{J_{1}}+1=0,\text{ for
}\frac{J_{2}}{J_{1}}\leq-\frac{1}{2}, \label{supn24}%
\end{equation}
where%
\begin{equation}
M_{2}^{z}=\frac{\arccos(\frac{J_{1}}{J_{2}}+1)}{\pi}-\frac{1}{2}.
\label{mznp2}%
\end{equation}

The critical lines, shown in Figs. \ref{fig01} and \ref{fig02}, can be
obtained from Eqs. \ref{sup21}-\ref{mzp2} for $\ J_{2}/J_{1}=0.2$ and
$J_{2}/J_{1}=2.0,$ and from Eqs. \ref{supn21}-\ref{mznp2} for $\ J_{2}%
/J_{1}=-0.2$ and $J_{2}/J_{1}=-2.0$.

It should be noted that for $J_{2}/J_{1}=1/2$ the critical surfaces meet at a
bicritical line\cite{Lenilson 2005} given by%
\begin{equation}
\frac{h}{J_{1}}=-\frac{3}{4}+\frac{I}{J_{1}}. \label{eqbiline}%
\end{equation}

For $I>0,$ where the phase transitions are of first-order, critical surfaces
are obtained numerically from the solution of the system%
\begin{equation}
\left\{
\begin{array}
[c]{c}%
M_{it}^{z}-\frac{\varphi_{2t}^{i}-\varphi_{1t}^{i}}{\pi}+\frac{1}{2}=0,\\
M_{jt}^{z}-\frac{\varphi_{2t}^{j}-\varphi_{1t}^{j}}{\pi}+\frac{1}{2}=0,\\
f(M_{it}^{z})-f(M_{jt}^{z})=0,
\end{array}
\right.  \label{sufirstp2}%
\end{equation}
where $\varphi_{2t}^{i}$ and $\varphi_{1t}^{i}$ are given by%

\begin{equation}
\varphi_{1t}^{i}=\arccos\left[  \frac{J_{1}+\sqrt{J_{1}^{2}+2J_{2}^{2}%
+4J_{2}\left(  h_{t}+2IM_{it}^{z}\right)  }}{2J_{2}}\right]  ,\text{ }
\label{phi1}%
\end{equation}

\begin{equation}
\varphi_{2t}^{i}=\arccos\left[  \frac{J_{1}-\sqrt{J_{1}^{2}+2J_{2}^{2}%
+4J_{2}\left(  h_{t}+2IM_{it}^{z}\right)  }}{2J_{2}}\right]  , \label{phi2}%
\end{equation}

\noindent$M_{it}^{z},$ $M_{jt}^{z}$ are the values of the magnetization at the
transition, with $i,$ $j=1,$ $2,$ $3,$ $4$ \ for $0\leq J_{2}/J_{1}\leq1/2$,
and with $i,$ $j=1,$ $2,$ $3,$ $4,$ $5,$ $6$ for $J_{2}/J_{1}\geq1/2,$ and
$f(M_{it}^{z})$ is given by Eq. (\ref{funquantum}). As in the case of second
order transitions, the critical lines shown in the Fig. \ref{fig01} are
determined from the previous systems by considering $J_{2}/J_{1}=0.2$ and
$J_{2}/J_{1}=2$.

For $J_{2}/J_{1}\geq1/2,$ there are four critical surfaces and three of them,
for $J_{2}/J_{1}=1/2,$ meet at a critical line which is determined by the
following system of equations%
\begin{equation}
\left\{
\begin{array}
[c]{c}%
M_{1t}^{z}-\frac{\varphi_{2t}^{1}-\varphi_{1t}^{1}}{\pi}+\frac{1}{2}=0,\\
M_{2t}^{z}-\frac{\varphi_{2t}^{2}-\varphi_{1t}^{2}}{\pi}+\frac{1}{2}=0,\\
M_{3t}^{z}-\frac{\varphi_{2t}^{3}-\varphi_{1t}^{3}}{\pi}+\frac{1}{2}=0,\\
f(M_{1t}^{z})-f(M_{3t}^{z})=0,\\
f(M_{2t}^{z})-f(M_{3t}^{z})=0,
\end{array}
\right.  \label{systemequations}%
\end{equation}
where $f\left(  M_{it}^{z}\right)  $ is obtained from the Eq.
(\ref{funquantum}), and with $i=1,$ $2,$ $3$ and $M_{3t}^{z}=-1/2.$ This is a
triple line which meets the bicritical line at $I=0.$

A second triple line can be determined by imposing $M_{3t}^{z}=-1/2$ and
$M_{4t}^{z}=1/2$ in the system
\begin{equation}
\left\{
\begin{array}
[c]{c}%
M_{2t}^{z}-\frac{\varphi_{2t}^{2}}{\pi}+\frac{1}{2}=0,\\
f(M_{2t}^{z})-f(M_{3t}^{z})=0,\\
f(M_{2t}^{z})-f(M_{4t}^{z})=0,
\end{array}
\right.  \label{systemequ2}%
\end{equation}
where $f(M_{2t}^{z}),$ and $\varphi_{2t}^{2}$ are given by Eqs.
(\ref{funquantum}) and (\ref{phi2}), which begins at the point $J_{2}%
/J_{1}=0,$ $h/J_{1}=0$ and $I/J_{1}=4/\pi,$ which has been obtained
Gon\c{c}alves et al. \cite{Lenilson 2005}$.$

\begin{figure}
[ptb]
\begin{center}
\includegraphics[
natheight=8.853900in,
natwidth=11.458800in,
height=4.3664in,
width=5.642in
]%
{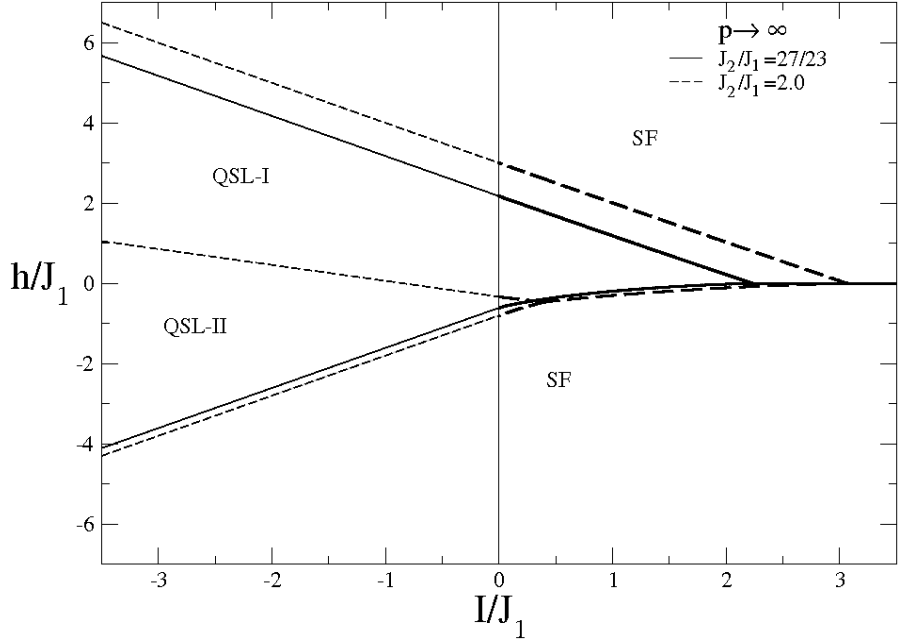}%
\caption{ Phase diagram for the quantum transitions \ as a function of the
long-range interaction $I/J_{1}$ for $p\rightarrow\infty$ and $J_{2}%
/J_{1}=27/23,$ $2.0.$ For $J_{2}/J_{1}=27/23$ there are three phases, one spin
liquid phase (QSL-I) and two saturated ferromagnetic phases(SF), and for
$J_{2}/J_{1}=2.0,$ there are four phases, two spin liquid phases
(QSL-I,QSL-II) and two saturated ferromagnetic phases(SF). The critical lines
correspond to first-order phase transitions for $I/J_{1}>0$ and to
second-order phase transitions for $I/J_{1}\leq0.$ }%
\label{fig07}%
\end{center}
\end{figure}

\begin{figure}
[ptb]
\begin{center}
\includegraphics[
natheight=8.853900in,
natwidth=11.458800in,
height=4.3664in,
width=5.642in
]%
{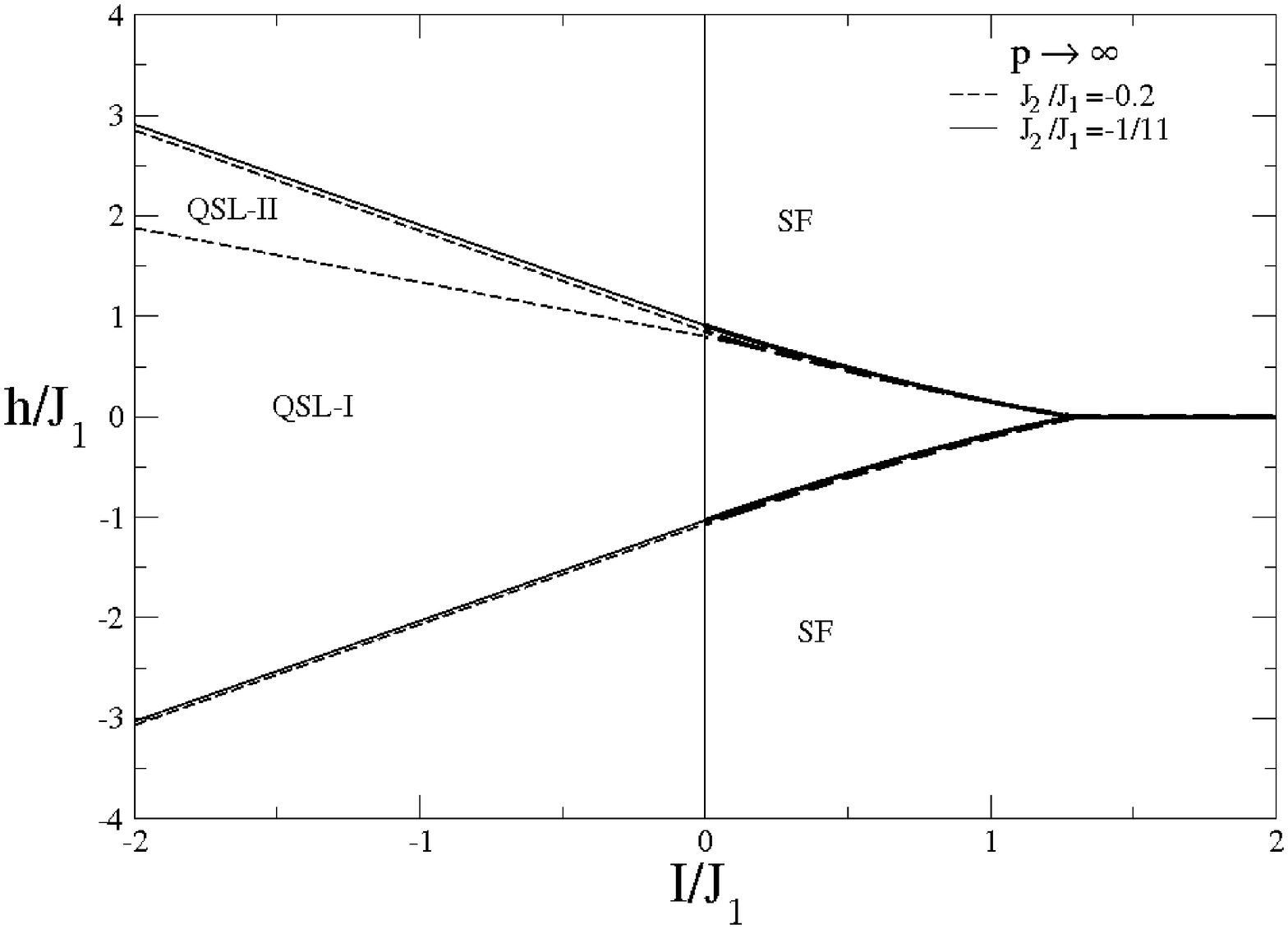}%
\caption{ Phase diagram for the quantum transitions \ as a function of the
long-range interaction $I/J_{1}$ for $p\rightarrow\infty$ and $J_{2}%
/J_{1}=-0.2,-1/11.$ For $J_{2}/J_{1}=-0.2$ there are four phases, two spin
liquid phases (QSL-I,QSL-II) and two saturated ferromagnetic phases(SF), and
for $J_{2}/J_{1}=-1/11,$ there are three phases, one spin liquid phase (QSL)
and two saturated ferromagnetic phases(SF). The critical lines correspond to
first-order phase transitions for $I/J_{1}>0$ and to second-order phase
transitions for $I/J_{1}\leq0.$ }%
\label{fig08}%
\end{center}
\end{figure}

\begin{figure}
[ptb]
\begin{center}
\includegraphics[
natheight=8.853900in,
natwidth=11.458800in,
height=4.3664in,
width=5.642in
]%
{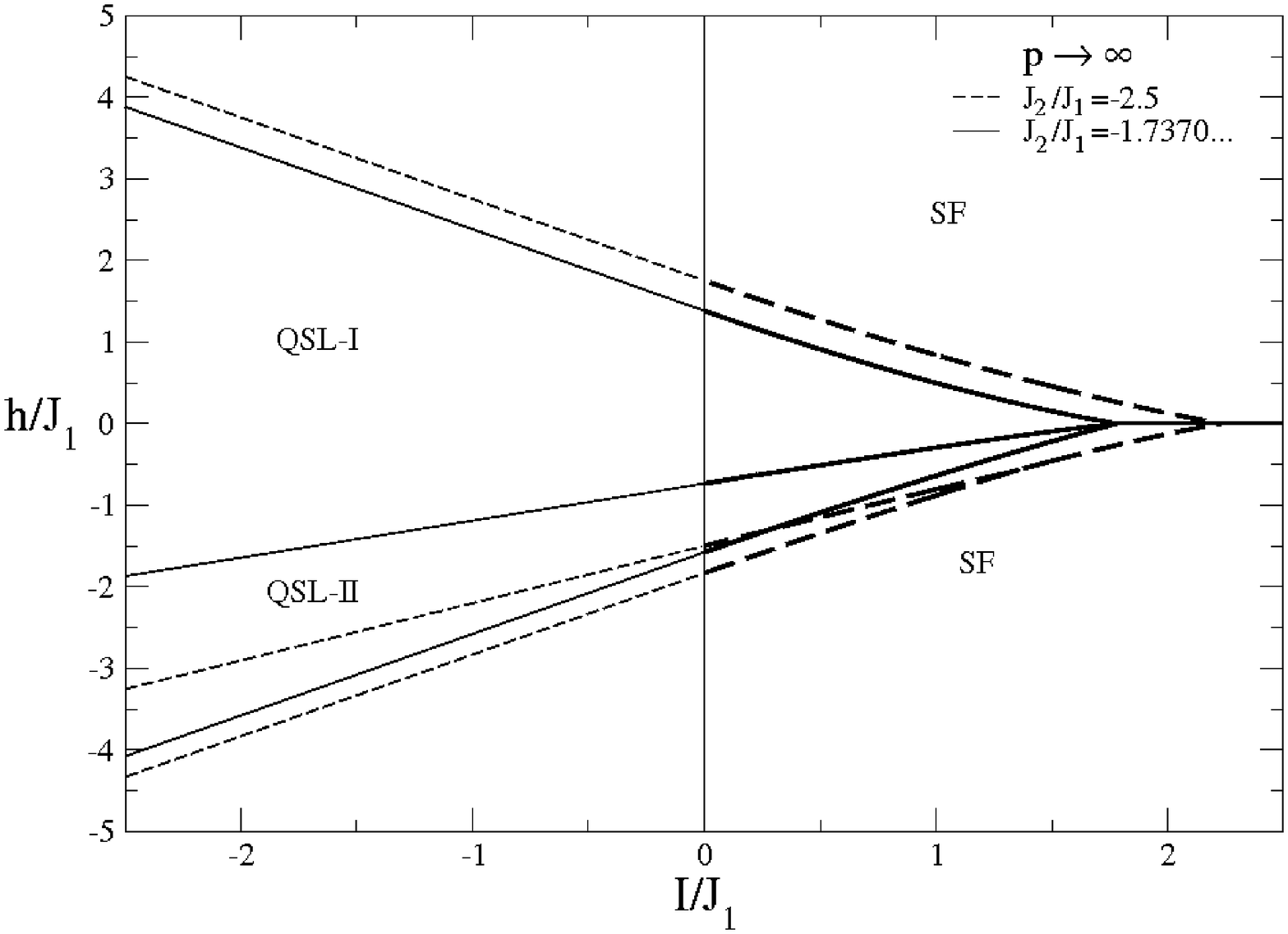}%
\caption{ Phase diagram for the quantum transitions \ as a function of the
long-range interaction $I/J_{1}$ for $p\rightarrow\infty$ and $J_{2}%
/J_{1}=-2.5,-1.7370...$ For $J_{2}/J_{1}=-2.5$ there are four phases, two spin
liquid phases (QSL-I,QSL-II) and two saturated ferromagnetic phases(SF), and
for $J_{2}/J_{1}=-1.7370..,$ there are four phases, two spin liquid phases
(QSL-I,QSL-II)) and two saturated ferromagnetic phases(SF) four phases which
coexist in the quadruple point localized at $h/J_{1}=0,$ $I/J_{1}=1.7798...$
The functional of the free energy for this case is shown in Fig. \ref{fig11}.
The critical lines correspond to first-order phase transitions for $I/J_{1}>0$
and to second-order phase transitions for $I/J_{1}\leq0.$ }%
\label{fig09}%
\end{center}
\end{figure}

\begin{figure}
[ptb]
\begin{center}
\includegraphics[
natheight=8.853900in,
natwidth=11.458800in,
height=4.3664in,
width=5.642in
]%
{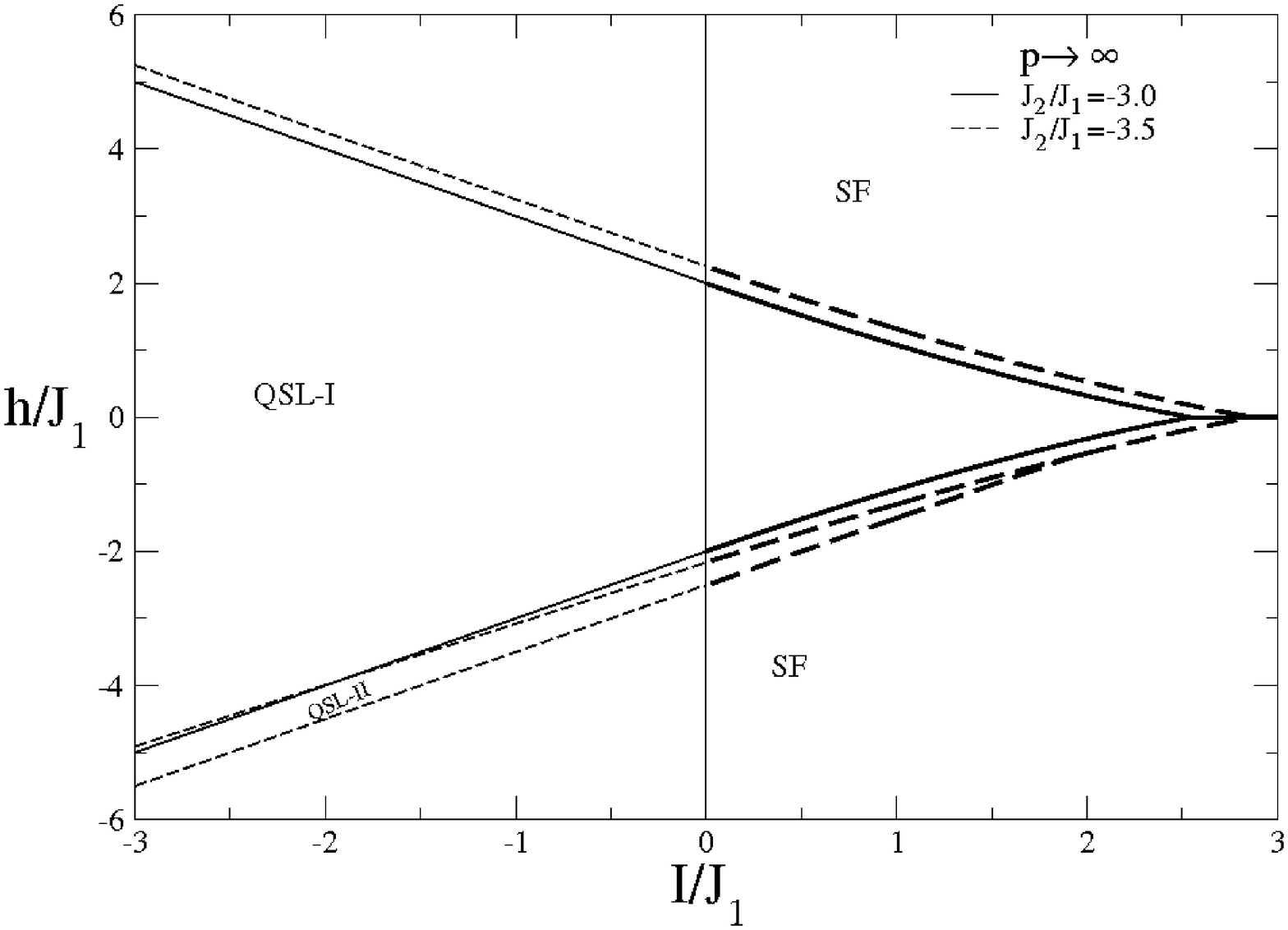}%
\caption{ Phase diagram for the quantum transitions \ as a function of the
long-range interaction $I/J_{1}$ for $p\rightarrow\infty$ and $J_{2}%
/J_{1}=-3.5,-3.0.$ For $J_{2}/J_{1}=-3.5$ there are four phases, two spin
liquid phases (QSL-I,QSL-II) and two saturated ferromagnetic phases(SF), and
for $J_{2}/J_{1}=-3.0,$ there are three phases, one spin liquid phase (QSL-I)
and two saturated ferromagnetic phases(SF).The critical lines correspond to
first-order phase transitions for $I/J_{1}>0$ and to second-order phase
transitions for $I/J_{1}\leq0.$ }%
\label{fig10}%
\end{center}
\end{figure}

For the special case $p\rightarrow\infty$, we can find the critical surfaces
by using the same procedure used in the case $p=2$. In this case, due to many
intersections of the critical surfaces, the global phase diagram becomes too
complicated, as we will show below. Therefore, we will present some
projections of the global diagram which contain the main characteristics of
this diagram and are shown in Figs. \ref{fig07}-\ref{fig10}.

In this case, the fermion excitation spectrum, obtained from Eq.
(\ref{energyequation}), is given by%
\begin{equation}
\epsilon_{q}\left(  M^{z}\right)  =J_{1}\cos(q)-J_{2}\left[  \frac
{2\cos\left(  2q\right)  +\cos\left(  q\right)  }{5+4\cos\left(  q\right)
}\right]  +h+2IM^{z}, \label{enerpinf}%
\end{equation}
and from this result we can \ determine the equations of the critical surfaces
for $I/J_{1}\leq0$ and $J_{2}/J_{1}\geq0,$ which are
\begin{equation}
\frac{J_{2}}{J_{1}}-\frac{I}{J_{1}}-\frac{h}{J_{1}}+1=0,\text{ for }%
\frac{J_{2}}{J_{1}}\geq0, \label{supinf1}%
\end{equation}%
\begin{equation}
\frac{J_{2}}{3J_{1}}+\frac{I}{J_{1}}-\frac{h}{J_{1}}-1=0,\text{ for }%
0\leq\frac{J_{2}}{J_{1}}\leq\frac{27}{23}, \label{supinf2}%
\end{equation}%
\begin{equation}
\frac{5J_{1}-9J_{2}}{4J_{1}}+\sqrt{\frac{3J_{2}}{J_{1}}\left(  \frac{J_{2}%
}{J_{1}}-1\right)  }+\frac{I}{J_{1}}-\frac{h}{J_{1}}=0,\text{ for }\frac
{J_{2}}{J_{1}}\geq\frac{27}{23}, \label{supinf3}%
\end{equation}%
\begin{equation}
\frac{J_{2}}{3J_{1}}-\frac{2IM_{1}^{z}}{J_{1}}-\frac{h}{J_{1}}-1=0,\text{ for
}\frac{J_{2}}{J_{1}}\geq\frac{27}{23}, \label{supinf4}%
\end{equation}
where%
\begin{equation}
M_{1}^{z}=\frac{\arccos\left(  \frac{11J_{2}-15J_{1}}{12J_{1}-12J_{2}}\right)
}{\pi}-\frac{1}{2}. \label{mz1pinf}%
\end{equation}

\bigskip Identically, we can show that for $I/J_{1}\leq0$ and $J_{2}/J_{1}%
\leq0$, the critical surfaces are%
\begin{equation}
\frac{J_{2}}{J_{1}}-\frac{I}{J_{1}}-\frac{h}{J_{1}}+1=0,\text{ for }-\frac
{1}{11}\leq\frac{J_{2}}{J_{1}}\leq0, \label{supinf5}%
\end{equation}%
\begin{equation}
\frac{J_{2}}{3J_{1}}+\frac{I}{J_{1}}-\frac{h}{J_{1}}-1=0,\text{ for }%
-3\leq\frac{J_{2}}{J_{1}}\leq0, \label{supinf6}%
\end{equation}%
\begin{equation}
\frac{5J_{1}-9J_{2}}{4J_{1}}-\sqrt{\frac{3J_{2}}{J_{1}}\left(  \frac{J_{2}%
}{J_{1}}-1\right)  }-\frac{I}{J_{1}}-\frac{h}{J_{1}}=0,\text{ for }\frac
{J_{2}}{J_{1}}\leq-\frac{1}{11}, \label{supinf7}%
\end{equation}%
\begin{equation}
\frac{J_{2}}{J_{1}}-\frac{2IM_{2}^{z}}{J_{1}}-\frac{h}{J_{1}}+1=0,\text{ for
}-3\leq\frac{J_{2}}{J_{1}}\leq-\frac{1}{11}, \label{supinf8}%
\end{equation}%
\begin{equation}
\frac{J_{2}}{J_{1}}+\frac{I}{J_{1}}-\frac{h}{J_{1}}+1=0,\text{ for }%
\frac{J_{2}}{J_{1}}\leq-3, \label{supinfa}%
\end{equation}%
\begin{equation}
\frac{J_{2}}{3J_{1}}-\frac{2IM_{3}^{z}}{J_{1}}-\frac{h}{J_{1}}-1=0,\text{ for
}\frac{J_{2}}{J_{1}}\leq-3, \label{supinfb}%
\end{equation}
where%
\begin{equation}
M_{2}^{z}=\frac{\arccos\left(  -\frac{7J_{2}+5J_{1}}{4J_{1}-4J_{2}}\right)
}{\pi}-\frac{1}{2}, \label{mz2pinfb}%
\end{equation}%
\begin{equation}
M_{3}^{z}=\frac{\arccos\left(  \frac{-11J_{2}+15J_{1}}{12J_{1}-12J_{2}%
}\right)  }{\pi}-\frac{1}{2}. \label{mz3pinf}%
\end{equation}

In this case there are three bicritical lines which are given by%
\begin{equation}
\frac{h}{J_{1}}=-\frac{14}{23}+\frac{I}{J_{1}},\text{ at }J_{2}/J_{1}%
=27/23,\label{biline1pinf}%
\end{equation}%
\begin{equation}
\frac{h}{J_{1}}=\frac{10}{11}-\frac{I}{J_{1}},\text{ at }J_{2}/J_{1}%
=-1/11,\label{bipinf1}%
\end{equation}%
\begin{equation}
\frac{h}{J_{1}}=-2+\frac{I}{J_{1}},\text{ at }J_{2}/J_{1}=-3.\label{bipinf2}%
\end{equation}

For $I/J_{1}>0,$ as for the case $p=2$, the first order transition surfaces
can be determined numerically and the triple lines are determined by following
the same procedure adopted for $p=2.$

In Fig. \ref{fig07}  the phase diagram is shown for the positive region
$J_{2}/J_{1}\geq0,$ where there are four critical surfaces which are given by
the Eqs (\ref{supinf1}-\ref{mz1pinf}). These critical surfaces meet at a
bicritical line, which is given by Eq. \ref{biline1pinf}.

\begin{figure}
[ptb]
\begin{center}
\includegraphics[
natheight=8.853900in,
natwidth=11.458800in,
height=4.3664in,
width=5.642in
]%
{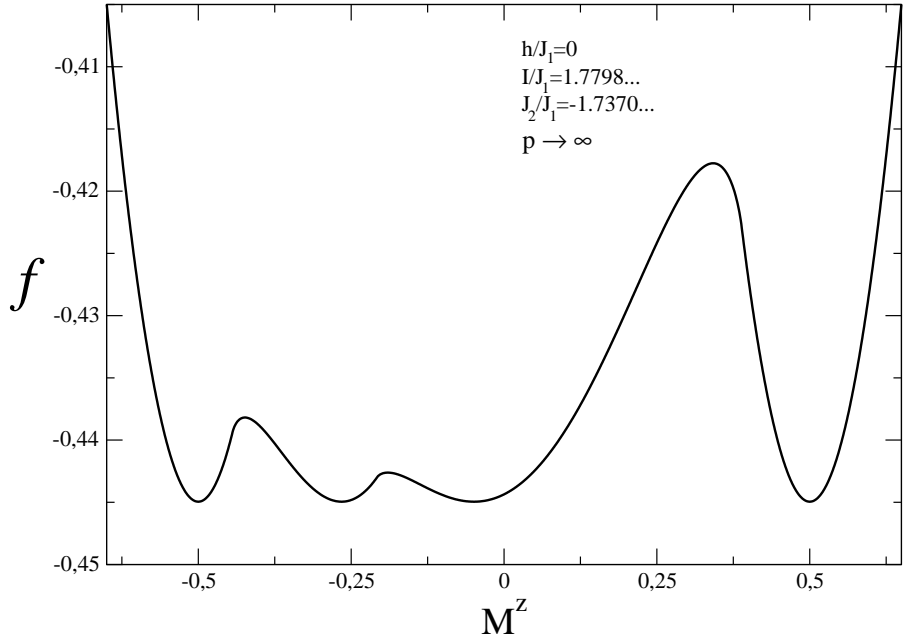}%
\caption{Functional of the free energy $f$ as a function of the magnetization
$M^{z}$ for $p\rightarrow\infty$, $h/J_{1}=0,$ $I/J_{1}=1.7798...$ and
$J_{2}/J_{1}=-1.7370...$ at the quadruple point.}%
\label{fig11}%
\end{center}
\end{figure}

The phase diagrams for the negative region $J_{2}/J_{1}\leq0$ and for
different values of $J_{2}/J_{1}$ are shown in Figs.\ref{fig08}-\ref{fig10}.
In this case there are six critical surfaces, given by Eqs \ref{supinf5}%
-\ref{supinfb}, which meet at bicritical lines given by the Eqs.(\ref{bipinf2}%
-\ref{bipinf1}). As we can see in Figs.\ref{fig08}-\ref{fig10}, the system
presents in this case identical critical behavior to the one obtained for the
case $p=2$, as far as the critical behavior is concerned. However, it is worth
mentioning  the appearance of quadruple point at $h/J_{1}=0,$ $J_{2}%
/J_{1}=-1.7370...,$ and $I/J_{1}=1.7798...,$which \ is shown in Fig.
\ref{fig09} and is not present in the case $p=2$. The behavior of the
functional of the Helmholtz free energy at this point is presented in Fig.
\ref{fig11}.

We have also analysed the cases $p=3$ and $p=4$, for $I=0$, where the model
presents new quantum spin liquid phases. Although in these cases there are no
first-order transitions, we have restricted these analyses to the model
without the long-range interaction, since the main purpose was to study
appearance of new quantum spin liquid phases, which is mainly controlled by
the multiple short-range interaction.

\begin{figure}
[ptb]
\begin{center}
\includegraphics[
natheight=8.853900in,
natwidth=11.458800in,
height=4.3664in,
width=5.642in
]%
{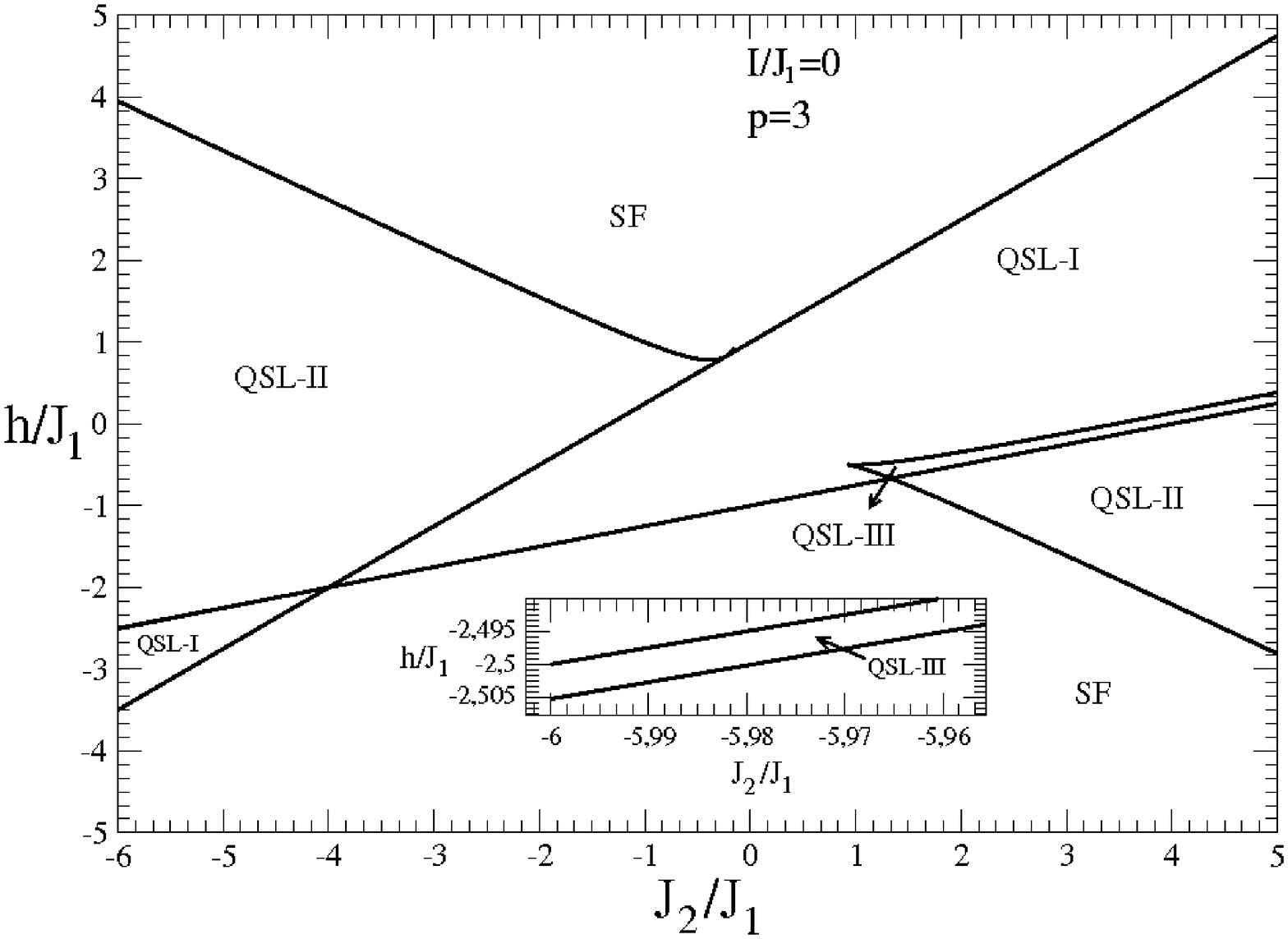}%
\caption{ Phase diagram for the quantum transitions as a function of
$J_{2}/J_{1}$ for $p=3$ and $I/J_{1}=0$. All the transitions are of
second-order and there are five phases, three spin liquid phase (QSL-I,
QSL-II, QSL-III) and two saturated ferromagnetic phases (SF).}%
\label{fig12}%
\end{center}
\end{figure}

In Fig. \ref{fig12} we show the phase diagram for the case $p=3,$ where the
critical lines are given by%
\begin{equation}
\frac{h}{J_{1}}=\frac{J_{2}}{4J_{1}}-1, \label{h1p3}%
\end{equation}%
\begin{equation}
\frac{h}{J_{1}}=\frac{3J_{2}}{4J_{1}}+1, \label{h2p3}%
\end{equation}%
\begin{equation}
\frac{h}{J_{1}}=\frac{1}{108}\left(  \frac{13J_{2}-12J_{1}}{J_{1}}\right)
^{\frac{3}{2}}-\frac{19J_{2}+36J_{1}}{108J_{1}},\text{ for }\frac{J_{2}}%
{J_{1}}\geq\frac{12}{13}\text{ or }\frac{J_{2}}{J_{1}}\leq-4, \label{h3p3}%
\end{equation}%
\begin{equation}
\frac{h}{J_{1}}=-\frac{1}{108}\left(  \frac{13J_{2}-12J_{1}}{J_{1}}\right)
^{\frac{3}{2}}-\frac{19J_{2}+36J_{1}}{108J_{1}},\text{for }\frac{J_{2}}{J_{1}%
}\geq\frac{12}{13}\text{ or }\frac{J_{2}}{J_{1}}<0. \label{h4p3}%
\end{equation}

As it can be seen, there are six quantum spin liquid phases which, as we will
show later, can be classified \ in three different classes as far as the
critical behavior is concerned.

\begin{figure}
[ptb]
\begin{center}
\includegraphics[
natheight=7.094100in,
natwidth=9.832900in,
height=4.2886in,
width=5.9343in
]%
{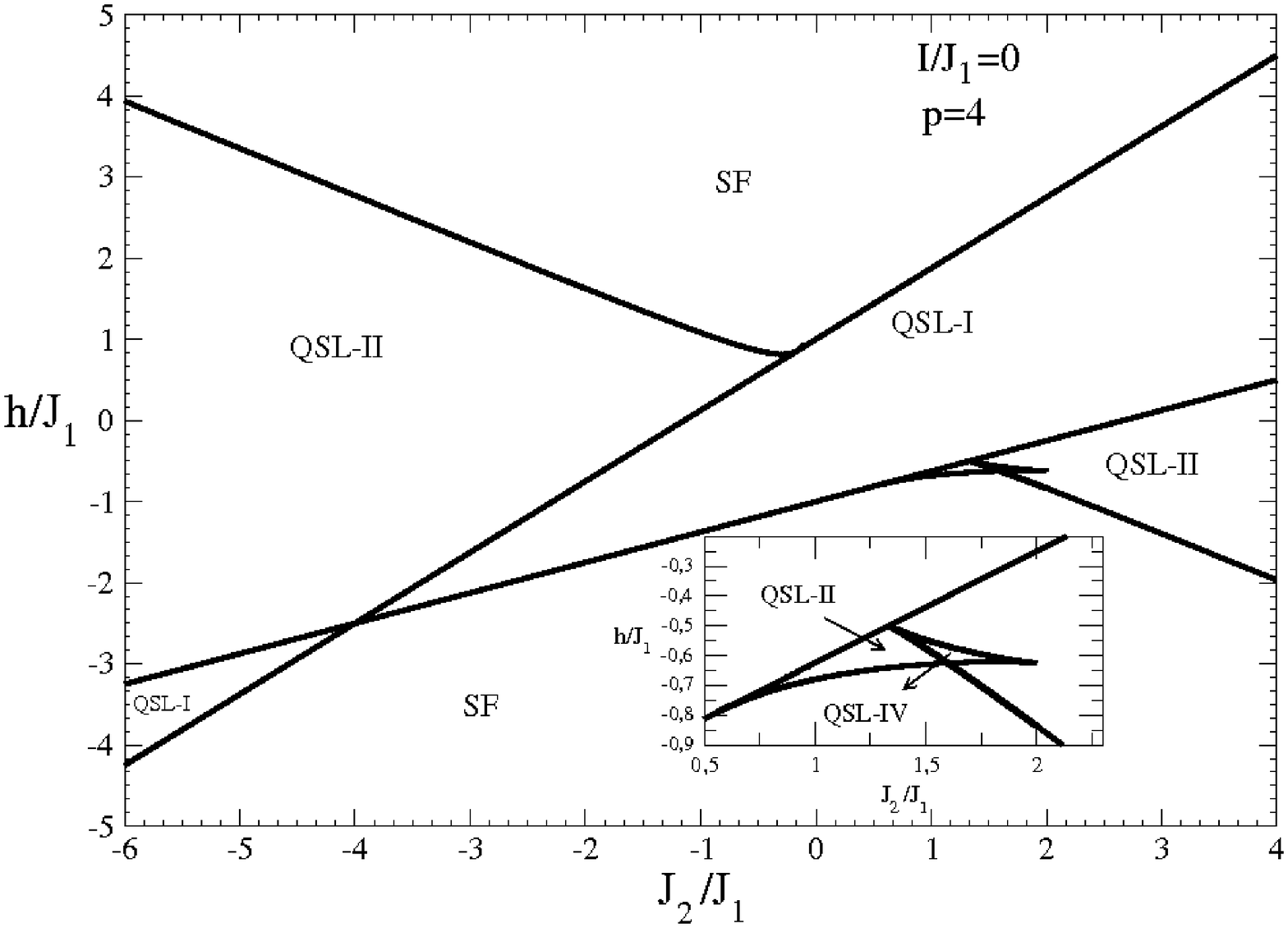}%
\caption{Phase diagram for the quantum transitions as a function of
$J_{2}/J_{1}$ for $p=4$ and $I/J_{1}=0$. All the transitions are of
second-order and there are five phases, three spin liquid phase (QSL-I,
QSL-II, QSL-IV) and two saturated ferromagnetic phases (SF).}%
\label{fig13}%
\end{center}
\end{figure}

The phase diagram for $p=4$ is shown in the Fig. \ref{fig13}, where the
critical lines are given by%
\begin{equation}
\frac{h}{J_{1}}=\frac{3J_{2}}{8J_{1}}-1, \label{h1p4}%
\end{equation}%
\begin{equation}
\frac{h}{J_{1}}=\frac{7J_{2}}{8J_{1}}+1, \label{h2p4}%
\end{equation}%
\begin{equation}
\frac{h}{J_{1}}=\frac{J_{2}}{16J_{1}}\left(  \frac{R_{1}}{2}\right)
^{4}-\frac{3J_{2}}{32J_{1}}\left(  \frac{R_{1}}{2}\right)  ^{2}+\left(
\frac{5J_{2}}{16J_{1}}-\frac{1}{2}\right)  \left(  \frac{R_{1}}{2}\right)
-\frac{51J_{2}}{256J_{1}}-\frac{1}{4}, \label{h3p4}%
\end{equation}%
\begin{equation}
\frac{h}{J_{1}}=\frac{J_{2}}{16J_{1}}\left(  \frac{R_{2}}{2}\right)
^{4}-\frac{3J_{2}}{32J_{1}}\left(  \frac{R_{2}}{2}\right)  ^{2}+\left(
\frac{1}{2}-\frac{5J_{2}}{16J_{1}}\right)  \left(  \frac{R_{2}}{2}\right)
-\frac{51J_{2}}{256J_{1}}-\frac{1}{4}, \label{h41p4}%
\end{equation}%
\begin{equation}
\frac{h}{J_{1}}=\frac{J_{2}R_{3i}^{4}}{J_{1}}-\frac{3J_{2}R_{3i}^{2}}{8J_{1}%
}+\left(  \frac{5J_{2}}{8J_{1}}-1\right)  R_{3i}-\left(  \frac{51J_{2}%
}{256J_{1}}-\frac{1}{4}\right)  , \label{hiip4}%
\end{equation}
where $R_{1}\equiv$ $K^{\frac{1}{3}}+K^{-\frac{1}{3}},$ with%

\begin{align}
K  &  \equiv(8J_{1}-5J_{2})/J_{2}+\sqrt{(64J_{1}^{2}-80J_{1}J_{2}+24J_{2}%
^{2})/J_{2}^{2}},\text{ \ \ }\nonumber\\
J^{\prime}  &  \equiv16b/(b^{2}+10b+1)\text{ and }b=\left[  (3+\sqrt
{5})/2\right]  ^{3}\text{ for }J^{\prime}\leq J_{2}/J_{1}\leq4/3,
\end{align}

$R_{2}\equiv S^{\frac{1}{3}}+S^{-\frac{1}{3}},$ with%

\begin{equation}
S\equiv(5J_{2}-8J_{1})/J_{2}+\sqrt{(24J_{2}^{2}-80J_{1}J_{2}+64J_{1}%
^{2})/J_{2}^{2}},\text{ for }J_{2}/J_{1}<0\text{ or }J_{2}/J_{1}\geq2,
\label{ss}%
\end{equation}

\noindent and $R_{3i}\equiv\cos\left(  \theta+\phi_{i}\right)  /2,$ with%

\begin{equation}
\theta=\arccos\left[  (8J_{1}-5J_{2})/J_{2}\right]  /3\text{ and }\phi
_{i}=2\pi\left(  i-1\right)  /3,\text{ }i=1,2,3\text{ for }4/3\leq J_{2}%
/J_{1}\leq2\text{.} \label{theta}%
\end{equation}

In this case, as for $p=3,$ the system presents six quantum spin liquid phases
which can also be classified in three classes \ as far as the critical
behavior is concerned, as well as there exist only second order phase
transitions.\textbf{ \bigskip}

\section{STATIC SPIN CORRELATIONS}

The static correlation function $\langle S_{j}^{z}S_{j+r}^{z}\rangle,$ in the
thermodynamic limit, can be given by\cite{siskens 1974,capel 1977,goncalves
1977},
\begin{equation}
\langle S_{j}^{z}S_{j+r}^{z}\rangle=\frac{Tr\left[  \exp(-\beta\mathcal{H}%
^{-})S_{j}^{z}S_{j+r}^{z}\right]  }{Tr\left[  \exp(-\beta\mathcal{H}%
^{-})\right]  }, \label{dynamicexp}%
\end{equation}
where $\mathcal{H}^{-}$ is the Hamiltonian given in the Eq. (\ref{decomposed}).

After the introduction of the Fourier transform[Eq. (\ref{fourier})],
$\mathcal{H}^{-}$ can be written in the form
\begin{equation}
\mathcal{H}^{-}=%
%TCIMACRO{\dsum \limits_{q}}%
%BeginExpansion
{\displaystyle\sum\limits_{q}}
%EndExpansion
\overline{\epsilon}_{q}\left(  M^{z}\right)  c_{q}^{\dagger}c_{q}+\frac{Nh}%
{2}, \label{hamilline fourier}%
\end{equation}
where $\overline{\epsilon}_{q}\left(  M^{z}\right)  $ is given by Eq.
(\ref{energyfinal}).

By introducing the fermion operators, we can write%
\begin{equation}
S_{j}^{z}=\frac{1}{N}%
%TCIMACRO{\dsum \limits_{qq^{\prime}}}%
%BeginExpansion
{\displaystyle\sum\limits_{qq^{\prime}}}
%EndExpansion
\exp[ij(q-q^{\prime})]c_{q}^{\dagger}c_{q}-\frac{1}{2},\label{szevolu}%
\end{equation}

\noindent and from Eq. (\ref{szevolu}), by using the Wick's
theorem\cite{Mattuck 1992}, the static correlation can be written as%
\begin{equation}
\langle S_{j}^{z}S_{j+r}^{z}\rangle=\left[  \frac{1}{N}%
%TCIMACRO{\dsum \limits_{q}}%
%BeginExpansion
{\displaystyle\sum\limits_{q}}
%EndExpansion
\langle n_{q}\rangle-\frac{1}{2}\right]  ^{2}+\frac{1}{N^{2}}%
%TCIMACRO{\dsum \limits_{qq^{\prime}}}%
%BeginExpansion
{\displaystyle\sum\limits_{qq^{\prime}}}
%EndExpansion
\exp\left[  i(q-q^{\prime})r\right]  (1-\langle n_{q^{\prime}}\rangle
),\label{statiwick}%
\end{equation}
with%
\begin{equation}
\langle n_{q}\rangle=\frac{1}{1+\exp\left[  \overline{\epsilon}_{q}\left(
M^{z}\right)  \right]  }.\label{ocupanumber}%
\end{equation}

From Eq. (\ref{statiwick}), we can obtain the static correlation function
which, after some straightforward calculations, can be written as%
\begin{equation}
\langle S_{j}^{z}S_{j+r}^{z}\rangle=\frac{1}{4}\left\{  \left[  \frac{1}{N}%
%TCIMACRO{\dsum \limits_{q}}%
%BeginExpansion
{\displaystyle\sum\limits_{q}}
%EndExpansion
\tanh\left(  \frac{\overline{\epsilon}_{q}}{2}\right)  \right]  ^{2}-\left[
\frac{1}{N}%
%TCIMACRO{\dsum \limits_{q}}%
%BeginExpansion
{\displaystyle\sum\limits_{q}}
%EndExpansion
\cos(qr)\tanh\left(  \frac{\overline{\epsilon}_{q}}{2}\right)  \right]
^{2}+\delta_{r0}\right\}  ,\label{sumstatic}%
\end{equation}
and, in the thermodynamic limit, in the form%
\begin{equation}
\langle S_{j}^{z}S_{j+r}^{z}\rangle=\left[  \frac{1}{2\pi}%
%TCIMACRO{\dint \limits_{0}^{\pi}}%
%BeginExpansion
{\displaystyle\int\limits_{0}^{\pi}}
%EndExpansion
\tanh\left(  \frac{\overline{\epsilon}_{q}}{2}\right)  dq\right]  ^{2}-\left[
\frac{1}{2\pi}%
%TCIMACRO{\dint \limits_{0}^{\pi}}%
%BeginExpansion
{\displaystyle\int\limits_{0}^{\pi}}
%EndExpansion
\cos(qr)\tanh\left(  \frac{\overline{\epsilon}_{q}}{2}\right)  dq\right]
^{2}+\frac{\delta_{r0}}{4}.\label{staticsz}%
\end{equation}

As it is well known\cite{McCoy 1971},\ at $T=0$, the direct longitudinal
correlation function of the short-range XY-model behaves asymptotically$\ $as
$\rho^{zz}(r)\equiv\langle S_{j}^{z}S_{j+r}^{z}\rangle-\left(  M^{z}\right)
^{2}\sim f(r)r^{-2},$ where $f(r)$ is an oscillatory function. A similar
behavior can be found for the above expression[Eq.\ref{staticsz}] for the
so-called quantum spin liquid phases, where the oscillatory factor $f(r)$
undergoes changes for different phases, as shown by Titvinidze and
Japaradze\cite{Titvinidze 2003}, for the case $p=2$ without long-range interaction.

In the presence of the long-range interaction, for $p=2$ and $I/J_{1}\leq0,$
we have at most two multiple quantum spin liquid phases in the limit
$T\rightarrow0.$ For the first one phase the parameters satisfy the conditions%

\begin{align}
0  &  \leq\frac{J_{2}}{J_{1}}\leq\frac{1}{2},\label{inter1szsz}\\
-1+\frac{J_{2}}{2J_{1}}+\frac{I}{J_{1}}  &  \leq\frac{h}{J_{1}}\leq
1+\frac{J_{2}}{2J_{1}}-\frac{I}{J_{1}},\nonumber
\end{align}
or \ %

\begin{align}
\frac{J_{2}}{J_{1}}  &  \geq\frac{1}{2},\label{inter2szsz}\\
-1+\frac{J_{2}}{2J_{1}}-\frac{2IM_{1}^{z}}{J_{1}}  &  <\frac{h}{J_{1}}%
\leq1+\frac{J_{2}}{2J_{1}}-\frac{I}{J_{1}},\nonumber
\end{align}
where $M_{1}^{z}$ is given by Eq. (\ref{mzp2}), and from Eq. (\ref{staticsz})
we obtain
\begin{equation}
\rho^{zz}(r)\sim-\frac{\sin^{2}[q_{1}r]}{\pi^{2}r^{2}}, \label{staszpa1}%
\end{equation}
where\
\begin{equation}
q_{1}=\arccos\left[  \frac{J_{1}-y\left(  M^{z}\right)  }{2J_{2}}\right]  ,
\label{kappa1}%
\end{equation}
with%
\begin{equation}
y\left(  M^{z}\right)  \equiv\sqrt{J_{1}^{2}+2J_{2}^{2}+4J_{2}\left(
h+2IM^{z}\right)  }. \label{lambda1}%
\end{equation}

For the second quantum spin liquid phase the parameters satisfy the conditions%

\begin{align}
\frac{J_{2}}{J_{1}}  &  \geq\frac{1}{2},\label{inter3szsz}\\
-\frac{J_{2}}{2J_{1}}-\frac{J_{1}}{4J_{2}}+\frac{I}{J_{1}}  &  \leq\frac
{h}{J_{1}}\leq-1+\frac{J_{2}}{2J_{1}}-\frac{2IM_{1}^{z}}{J_{1}},\nonumber
\end{align}
and, in this case, the asymptotic behavior of the direct correlation
$\rho^{zz}(r)$ is given by%

\begin{equation}
\rho^{zz}(r)\sim-\frac{\left\{  \sin[q_{1}r]-\sin[q_{2}r]\right\}  ^{2}}%
{\pi^{2}r^{2}}, \label{staszpa2}%
\end{equation}
where $q_{1}$ is given by Eq. (\ref{kappa1}) and $q_{2}$ by
\begin{equation}
q_{2}=\arccos\left[  \frac{J_{1}+y\left(  M^{z}\right)  }{2J_{2}}\right]  .
\label{kappa2}%
\end{equation}

\noindent Therefore, $\rho^{zz}(r)$ \ shows a similar behavior to the case
where the model does not contain long-range interaction\cite{Titvinidze 2003}.

The correlation length which diverges at the critical point\cite{Stanley
1971}, when we have a single phase, is associated to the period of the
oscillation of $\rho^{zz}(r)$ and is defined by an analytical extension of its
scaling form given by\cite{Lima 1994}%

\begin{equation}
\rho^{zz}(n)\text{ }\sim\frac{F(in/\xi)}{n^{p}}. \label{exana}%
\end{equation}
\ \ \ \ \

When the system presents two phases between the saturated ferromagnetic
phases, besides the correlation lengths $\xi_{1}$ there is an aditional one
$\xi_{2},$ associated to the adjacent transitions and, in this case, we can
define a further extension of the scaling form as%

\begin{equation}
\rho^{zz}(n)\sim\frac{F(in/\xi_{1})+F(in/\xi_{2})}{n^{p}}, \label{exana1}%
\end{equation}
and define lenght of the system $\xi$ in the form%

\begin{equation}
\xi=\xi_{1}+\xi_{2}. \label{lengcorr1}%
\end{equation}

Therefore, for the case where we have a single phase, by using Eq.
(\ref{exana}), we can find from Eq. (\ref{staszpa1})%

\begin{equation}
\frac{1}{\xi}=2\arccos\left[  \left\vert \frac{J_{1}-y\left(  M^{z}\right)
}{2J_{2}}\right\vert \right]  , \label{lengcorr}%
\end{equation}
which diverges at the critical points given by $h/J_{1}=1+J_{2}/2J_{1}%
-I/J_{1}$ and $h/J_{1}=-1+J_{2}/2J_{1}+I/J_{1}.$

On the other hand, for the case where we have two phases between the saturated
ferromagnetic phases, by using Eqs. (\ref{exana1}), (\ref{lengcorr1}) and, by
using Eq. (\ref{staszpa2}) we can find%

\begin{equation}
\frac{1}{\xi}=\frac{\arccos\left[  \left\vert \frac{J_{1}+y\left(
M^{z}\right)  }{2J_{2}}\right\vert \right]  +\arccos\left[  \left\vert
\frac{J_{1}-y\left(  M^{z}\right)  }{2J_{2}}\right\vert \right]  }%
{2\arccos\left[  \left\vert \frac{J_{1}-y\left(  M^{z}\right)  }{2J_{2}%
}\right\vert \right]  \left\{  \arccos\left[  \left\vert \frac{J_{1}+y\left(
M^{z}\right)  }{2J_{2}}\right\vert \right]  -\arccos\left[  \left\vert
\frac{J_{1}-y\left(  M^{z}\right)  }{2J_{2}}\right\vert \right]  \right\}  },
\label{lengcorr2}%
\end{equation}
which, as expected, diverges at the critical points $h/J_{1}=-1+J_{2}%
/2J_{1}-2IM_{1}^{z}/J_{1},$ $h/J_{1}=1+J_{2}/2J_{1}-I/J_{1}$ and
$h/J_{1}=-J_{2}/2J_{1}-J_{1}/4J_{2}+I/J_{1},$ where $M^{z}$ is an intermidiate
magnetization$.$

Although the results presented are for $p=2,$ it can be shown that the
correlation lenght always diverges at all second order critical points
irrespective of the value of $p,$ and for multiple transitions an aditional
correlation length has to be introduced for each new phase presented by the
system between the saturated ferromagnetic phases. Consequently, the scaling
relation, given in Eq.(\ref{exana1}), and teh correlation length of the system
have to be redefined accordingly.

The transversal correlation function $\langle S_{j}^{x}S_{j+r}^{x}\rangle,$
given by the Toeplitz determinant\cite{Lieb 1961}
\begin{equation}
\langle S_{j}^{x}S_{j+r}^{x}\rangle=\frac{1}{4}\left(
\begin{array}
[c]{ccccc}%
\langle A_{1}B_{2}\rangle & \langle A_{1}B_{3}\rangle & \langle A_{1}%
B_{4}\rangle & ... & \langle A_{1}B_{j+r+1}\rangle\\
\langle A_{2}B_{2}\rangle & \langle A_{2}B_{3}\rangle & \langle A_{2}%
B_{4}\rangle & ... & \langle A_{2}B_{j+r+1}\rangle\\
\langle A_{3}B_{2}\rangle & \langle A_{3}B_{3}\rangle & \langle A_{3}%
B_{4}\rangle & ... & \langle A_{3}B_{j+r+1}\rangle\\
\vdots & \vdots & \vdots & \ddots & \vdots\\
\langle A_{j}B_{j+r+2}\rangle & \langle A_{j}B_{l+3}\rangle & \langle
A_{j}B_{l+4}\rangle & ... & \langle A_{j}B_{j+r+1}\rangle
\end{array}
\right)  , \label{determiant}%
\end{equation}
where%
\begin{equation}
\langle A_{j}B_{l}\rangle=-\frac{1}{\pi}%
%TCIMACRO{\dint \limits_{0}^{\pi}}%
%BeginExpansion
{\displaystyle\int\limits_{0}^{\pi}}
%EndExpansion
\cos[q(j-l)]\tanh\left(  \frac{\overline{\epsilon}_{q}\left(  M^{z}\right)
}{2}\right)  dq, \label{contrac}%
\end{equation}
can be evaluated numerically$.$ This correlation, for the usual short-range
XY-model and at $T=0$, behaves asymptotically$\ $as $\langle S_{j}^{x}%
S_{j+r}^{x}\rangle\sim r^{-\frac{1}{2}}$\cite{McCoy 1971}$.$ As in the case of
the longitudinal correlation, the asymptotic behavior of the
transversal\ correlation function $\langle S_{j}^{x}S_{j+r}^{x}\rangle$ also
presents changes in its power law decay due to the presence of the multiple
spin interaction. This result has been shown by Titvinidze and
Japaradze\cite{Titvinidze 2003}, in the case $p=2$ without long-range
interaction, which classified these phases as spin liquid phases.

\begin{figure}
[ptb]
\begin{center}
\includegraphics[
natheight=8.853900in,
natwidth=11.458800in,
height=4.2869in,
width=5.54in
]%
{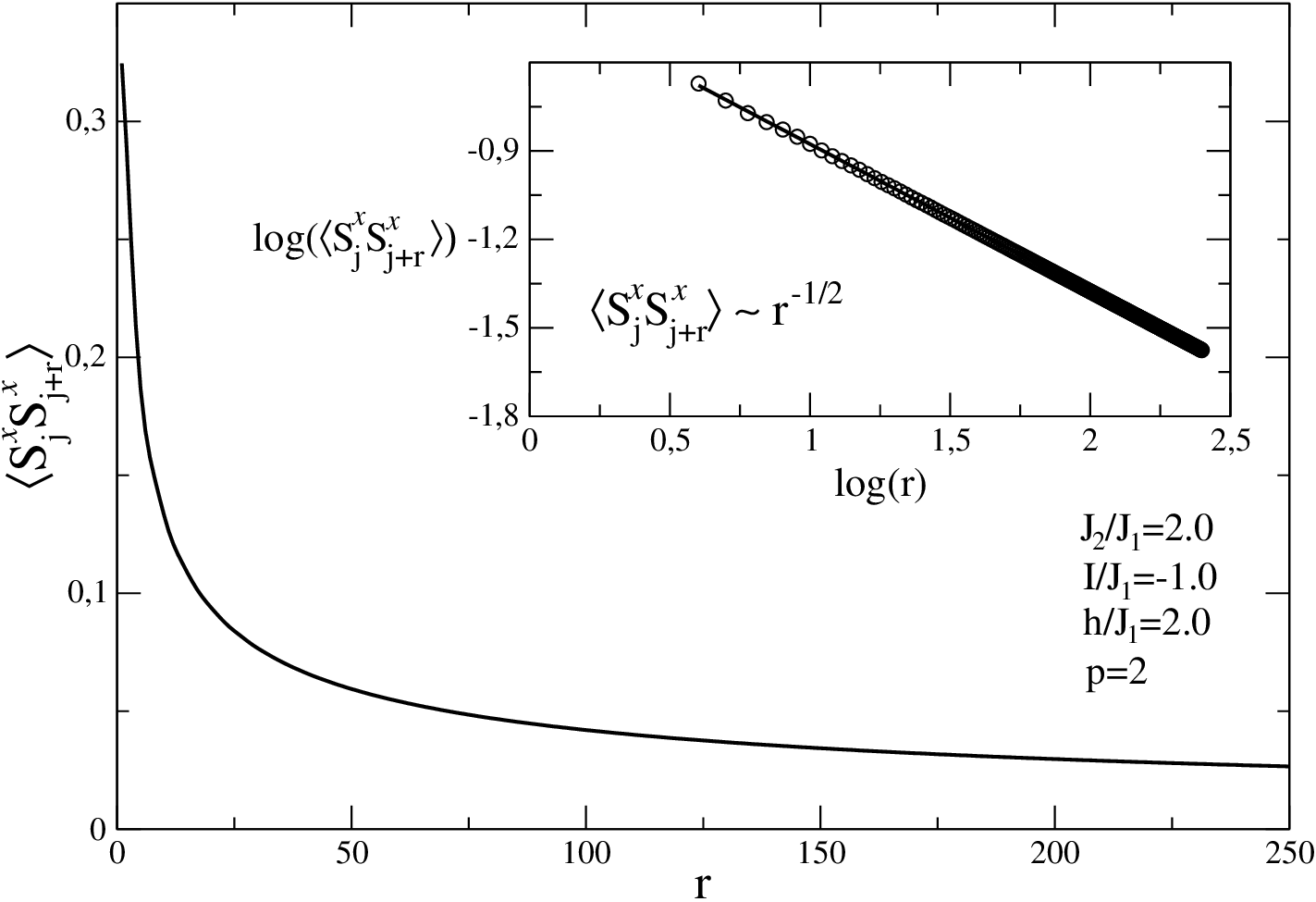}%
\caption{Static correlation function $\langle S_{j}^{x}S_{j+r}^{x}\rangle$as a
function of $r$, for $p=2$, $h/J_{1}=2.0,$ $J_{2}/J_{1}=2.0$ and
$I/J_{1}=-1.0.$ The inset shows the power law decay of the correlation
characteristic of the QSL-I phase.}%
\label{fig14}%
\end{center}
\end{figure}

\begin{figure}
[ptb]
\begin{center}
\includegraphics[
natheight=8.853900in,
natwidth=11.458800in,
height=4.2869in,
width=5.54in
]%
{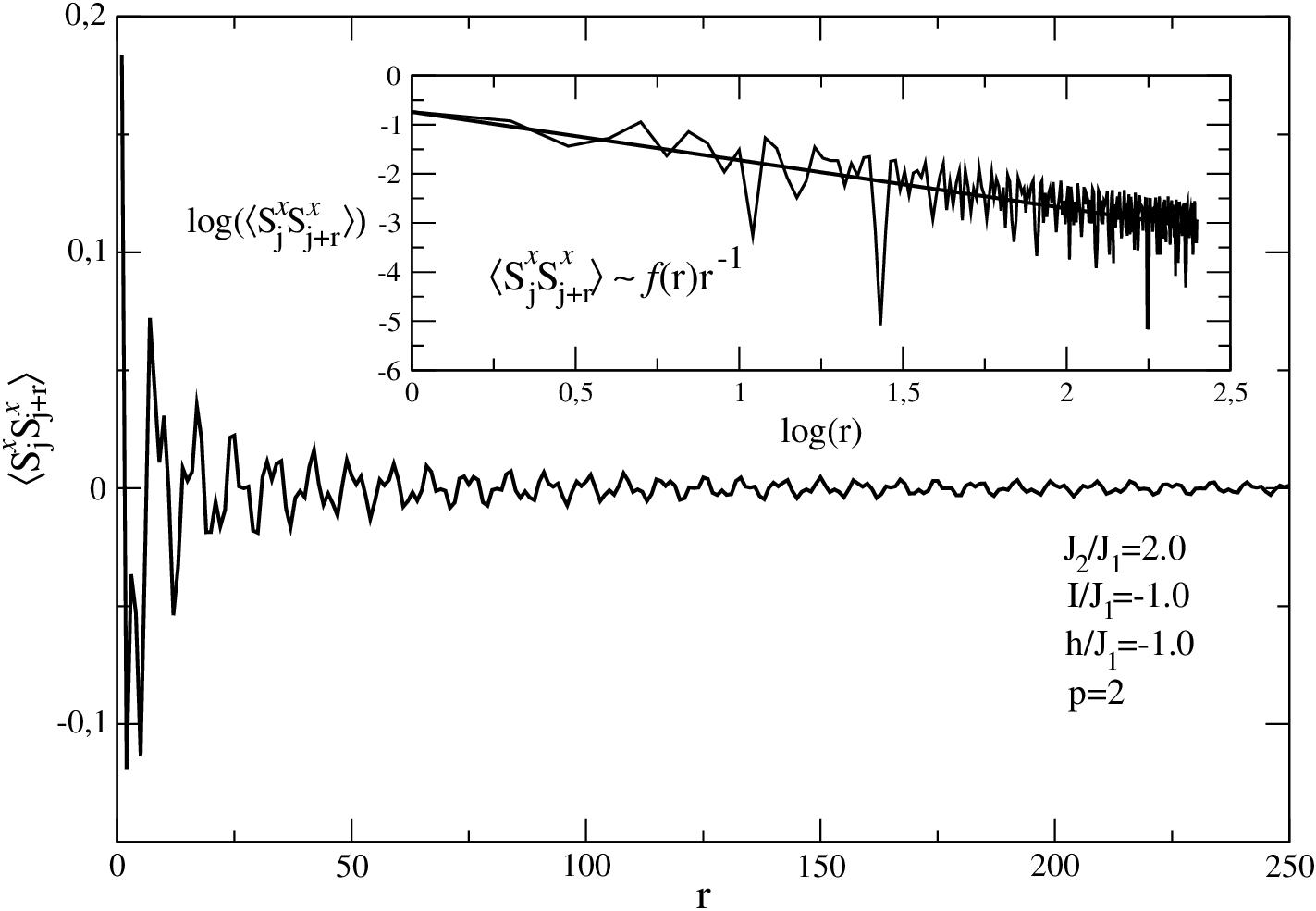}%
\caption{Static correlation function $\langle S_{j}^{x}S_{j+r}^{x}\rangle$as a
function of $r$, for $p=2$, $h/J_{1}=-1.0,$ $J_{2}/J_{1}=2.0$ and
$I/J_{1}=-1.0.$ The inset shows the power law decay of the correlation
characteristic of QSL-II phase with the oscillatory modulation\textbf{
}$f(r).$\textbf{ }}%
\label{fig15}%
\end{center}
\end{figure}

For the model with long-range interaction, the transversal static correlation,
at $T=0$, was evaluated numerically from Eq. (\ref{determiant}) by considering
the maximum value of $r$ equal to $250,$ and the power decay determined by
considering the scaling form $\langle S_{j}^{x}S_{l}^{x}\rangle\sim
f(r)r^{-\overline{p}}$. The results for $p=2,$ are presented in Figs.
\ref{fig14} and \ref{fig15} where the scaling forms are shown in the insets.
The transversal correlation function for the quantum spin liquid I phase,
presented in Fig. \ref{fig14}, the scaling behavior is given by $\langle
S_{j}^{x}S_{j+r}^{x}\rangle\sim r^{-\frac{1}{2}},$ while for the quantum spin
liquid II phase, presented in Fig. \ref{fig15}, it behaves as $\langle
S_{j}^{x}S_{j+r}^{x}\rangle\sim f(r)r^{-1},$ where $f(r)$ is an oscillatory
factor. These behaviors are identical to the ones obtained for the model
without long-range interaction\cite{Titvinidze 2003}.We would also like to
point out that\textbf{ }identical\textbf{ }results are obtained for the case
$p\rightarrow\infty.$ This general result, which depends on the number of
phases only, gives support to the classification of the intermediate phases as
spin liquid phases.

\begin{figure}
[ptb]
\begin{center}
\includegraphics[
natheight=8.853900in,
natwidth=11.458800in,
height=4.2869in,
width=5.54in
]%
{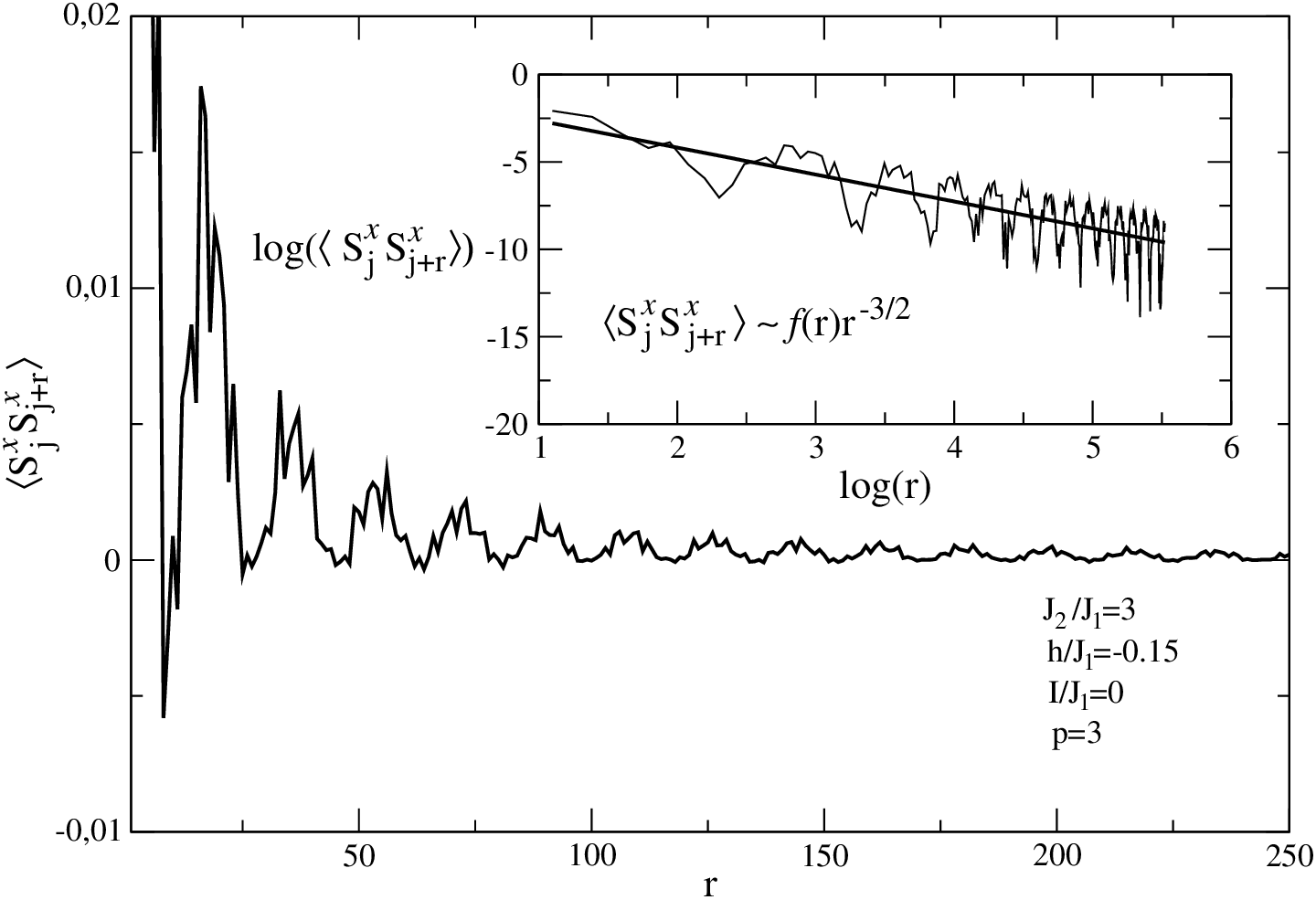}%
\caption{Static correlation function $\langle S_{j}^{x}S_{j+r}^{x}\rangle$as a
function of $r$, for $p=3$, $h/J_{1}=-0.15,$ $J_{2}/J_{1}=3.0$ and
$I/J_{1}=0.$ The inset shows the power law decay of the correlation
characteristic of QSL-III phase with the oscillatory modulation \textbf{
}$f(r).$}%
\label{fig16}%
\end{center}
\end{figure}

\begin{figure}
[ptb]
\begin{center}
\includegraphics[
natheight=8.853900in,
natwidth=11.458800in,
height=4.2869in,
width=5.54in
]%
{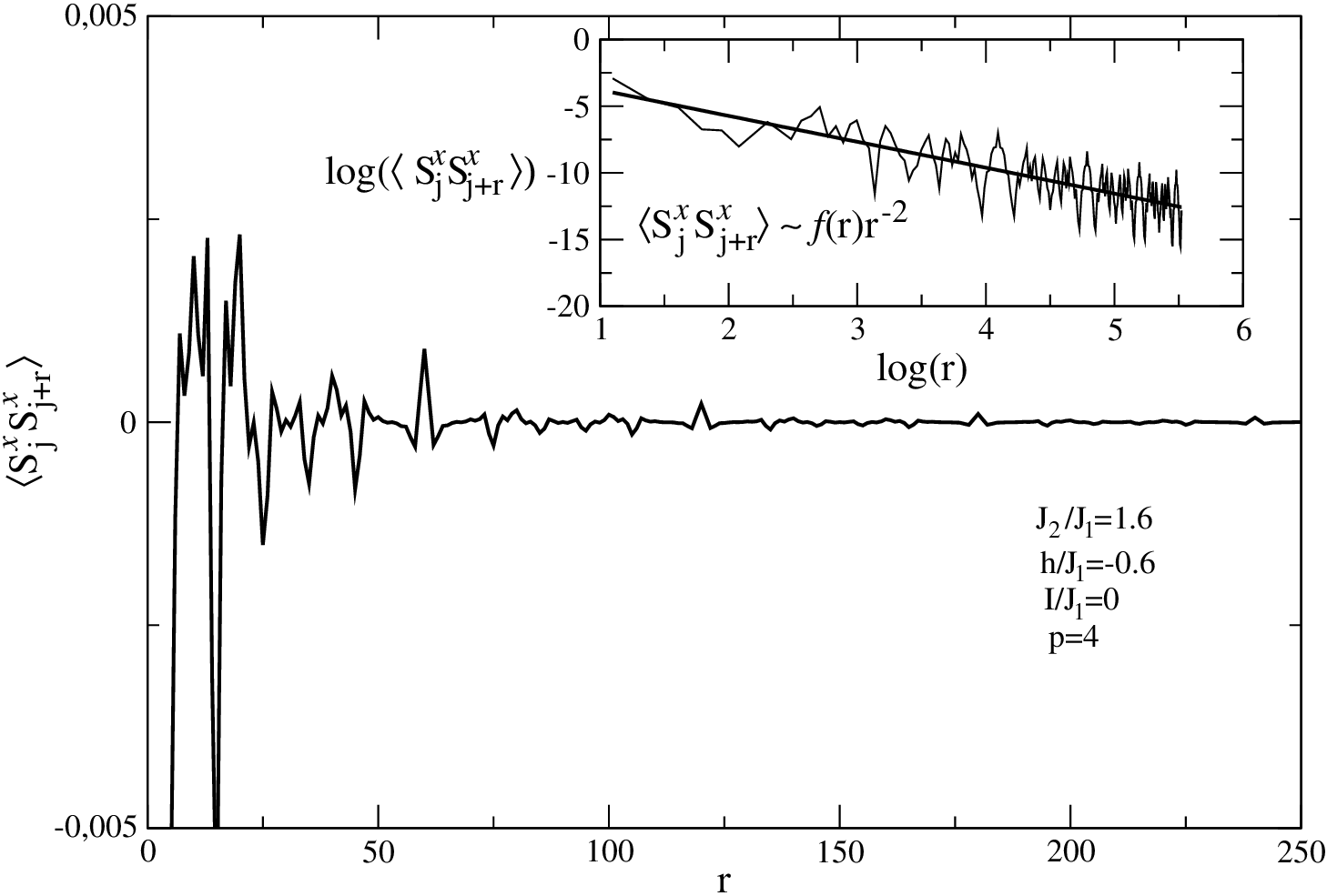}%
\caption{Static correlation function $\langle S_{j}^{x}S_{j+r}^{x}\rangle$as a
function of $r$, for $p=4$, $h/J_{1}=-0.6,$ $J_{2}/J_{1}=1.6$ and $I/J_{1}=0.$
The inset shows the power law decay of the correlation characteristic of
QSL-IV phase with the oscillatory modulation\textbf{ }$f(r).$\textbf{ }}%
\label{fig17}%
\end{center}
\end{figure}

The transversal static correlation was also calculated for $p=3,$ where there
is a new quantum spin liquid phase, which can be identified in the phase
diagram shown in Fig. \ref{fig12}. In this new phase, denominated quantum spin
liquid III phase, its scaling behavior is given by $\langle S_{j}^{x}%
S_{j+r}^{x}\rangle\sim f(r)r^{-\frac{3}{2}},$ where the oscillatory behavior
$f(r)$ is shown in Fig. \ref{fig16}. Finally, for $p=4,$ in the new phase
denominated quantum spin liquid IV phase and shown in the phase diagram
presented in Fig. \ref{fig13}, the scaling behavior of the transversal
correlation function is given by $\langle S_{j}^{x}S_{j+r}^{x}\rangle\sim
f(r)r^{-2}$, which is presented in Fig. \ref{fig17}.

It should be noted that, differently from the transversal correlation
function, the spatial decay of the longitudinal correlation function does not
depend on $\ p$.

\section{CRITICAL EXPONENTS}

The critical exponents, at $T=0,$ associated with the magnetization,
isothermal susceptibility, correlation length and the dynamic critical
exponent $z,$ for $p=2$, can be evaluated analytically since the quantities of
interest are known in closed form. Since these exponents are associated to
second order transitions, our analysis will be initially restricted to
$I\leq0.$

Therefore, following \cite{Lima 1994}, we define the order parameter given by
\begin{equation}
\widetilde{M^{z}}\equiv M_{t}^{z}-M^{z},\label{orderpara}%
\end{equation}
where $M_{t}^{z}$ is the magnetization at the transition. This order parameter
goes to zero at the second order transitions, and it is different from the one
proposed by Titvinidze and Japaridze\cite{Titvinidze 2003}. Then, from Eqs.
(\ref{magquantum}), (\ref{kappa1}) and (\ref{lambda1}), we get
\begin{equation}
M^{z}=\frac{1}{\pi}\arccos\left[  \frac{J_{1}-y\left(  M^{z}\right)  }{2J_{2}%
}\right]  -\frac{1}{2},\label{eqmzpara}%
\end{equation}
and by expanding Eq. (\ref{eqmzpara}) up to second-order in $\widetilde{M^{z}%
}$, we obtain%
\begin{equation}
\frac{\pi^{2}}{2}\left(  \widetilde{M^{z}}\right)  ^{2}\cong\frac{J_{1}\left(
h_{c}-h\right)  }{J_{1}+2J_{2}},\text{ \ \ \ for \ }\frac{I}{J_{1}%
}=0,\label{critieq}%
\end{equation}
and%
\begin{equation}
\widetilde{M^{z}}\cong-\frac{J_{1}\left(  h_{c}-h\right)  }{2I},\text{ \ \ for
\ }\frac{I}{J_{1}}<0.\label{critieq1}%
\end{equation}

From the Eqs. (\ref{critieq}) and (\ref{critieq1}), and the scaling form
$\widetilde{M^{z}}\sim\mid h_{c}-h$ $\mid^{\beta},$ we can conclude\ that the
critical exponent $\beta$ is given by $\beta=\frac{1}{2},$ when $I/J_{1}=0$
and $\beta=1$ for $I/J_{1}<0$, respectively, showing \ that the universality
class has changed with the presence of the long-range interaction.

The isothermal susceptibility can be obtained from the Eq. (\ref{eqmzpara}),
and is given by%
\begin{equation}
\chi_{T}^{zz}=\frac{J_{1}}{\pi\sqrt{\left[  y\left(  M^{z}\right)  \right]
^{2}\left[  1-\left(  \frac{J_{1}-y\left(  M^{z}\right)  }{2J_{2}}\right)
^{2}\right]  }-2I}.\label{xzexp}%
\end{equation}
In the critical region, we have \ $\chi_{T}^{zz}\sim\mid h_{c}-h$
$\mid^{\gamma},$ and from the previous expression we find that the critical
exponent $\gamma$ \ is \ equal to $\frac{1}{2},$ when $I/J_{1}=0,$ and \ it is
zero for $I/J_{1}<0$. Since at $T=0,$ \ $\gamma$ \ is identical to $\alpha,$
we can show that in both cases, namely, with or without long-range interaction
\ the exponents \ $\beta,$ $\alpha$ and $\gamma$ \ satisfy the Rushbrook
scaling relation\cite{Stanley 1971}.

The critical exponent $\nu$ associated with the correlation length $\xi
\sim\mid h_{c}-h\mid^{\nu}$, can be obtained from Eqs. (\ref{lengcorr}) and
(\ref{lengcorr2}). From these expressions we can immediately show that $\nu$
is \ equal to $\frac{1}{2},$ when $I/J_{1}=0$ and \ it is $1$ for $I/J_{1}<0$.

From the above results, by assuming the quantum hyperscaling
relation\cite{Continentino 2001}%
\begin{equation}
2-\alpha=\nu\left(  d+z\right)  , \label{hypsacal}%
\end{equation}
where $d$ is the dimension of the system and $z$ the dynamic critical
exponent, \ we find \ $z=1$ for $I<0,$ and $\ z=2$ for $I=0.$ Therefore, we
can conclude that the system also presents a non-universal critical dynamical behavior.

For $I>0,$ following Continentino and Ferreira\cite{Continentino 2004}, we
introduce a critical exponent associated to the free energy for the quantum
first-order phase transition. Assuming the scaling form $f(h^{\pm}%
)=f(h_{t})\pm E^{\pm}\mid h_{t}-h\mid^{2-\alpha}$ close to the field of
transition $h_{t},$\ since the free energy given in the Eq. (\ref{funquantum})
can be written as
\begin{equation}
f(\frac{h^{-}}{J_{1}})\cong f(\frac{h_{t}}{J_{1}})+M_{t}\mid\frac{h-h_{t}%
}{J_{1}}\mid,\text{ \ \ for \ }h<h_{t},\label{safun1}%
\end{equation}%
\begin{equation}
f(\frac{h^{+}}{J_{1}})\cong f(\frac{h_{t}}{J_{1}})-\frac{1}{2}\mid
\frac{h-h_{t}}{J_{1}}\mid,\text{ \ \ for \ }h>h_{t},\label{safun2}%
\end{equation}
we obtain $\alpha=1,$ which gives support to Continentino and Ferreira
conjecture\cite{Continentino 2004}$.$ \

\section{CONCLUSIONS}

In this work we have considered the one-dimensional isotropic XY model with
multiple spin interactions and uniform long-range interactions among the
\textit{z} components of the spin, in a transverse magnetic field. The
solution of the model was obtained exactly for arbitrary $p,$ which
characterizes the range of multiple spin interaction, and at arbitrary
temperature. Explicit equations have been obtained for the functional of the
Helmholtz free energy from which the equation of state can be determined
numerically. The quantum critical behavior was studied in details for
$p=2,3,4$ and $\infty,$ and the multiple phases presented by the model have
been characterized by the asymptotic behavior of the static correlation
\ functions $\langle S_{j}^{z}S_{l}^{z}\rangle$ and $\langle S_{j}^{x}%
S_{l}^{x}\rangle.$ Irrespective of the value of $p$, the model presents
first-order quantum transitions, when the long-range interaction is
ferromagnetic, and second-order ones, when the long-range interaction is
antiferromagnetic. Following \ Titvinidze and Japaridze\cite{Titvinidze 2003},
we have classified the intermediate phases, situated between saturated
feromagnetic phases, as the so-called quantum spin liquid phases, which are
induced by the extended interaction. For $p=2,$ the global phase diagram has
been obtained as a function of the interaction parameters and the critical
surfaces and multicritical lines determined exactly. The critical exponents
have been obtained and it has been verified that they satisfy the Rushbrook
relation $\alpha+$ 2$\beta+$ $\gamma=2$ (see e.g. ref.\cite{Stanley 1971}) and
the quantum hyperscaling relation\cite{Continentino 2001} $2-\alpha=\nu\left(
d+z\right)  ,$ and that the system presents a non-universal static and dynamic
critical behavior. We have also shown that the free energy, close to first
order transitions, satisfy the scaling form proposed by Continentino and
Ferreira\cite{Continentino 2004}.

Finally, we would like to point out that it is of great importance the
existence of multiple first order quantum transitions and multicritical lines
(triple and quadruple lines) , which we have shown exactly to exist in the
model, since we believe that they are related to the different mechanisms from
which the first order phase transitions are driven, as discusssed by
Pfleiderer\cite{Pfleiderer 2005}.

\section{ACKNOWLEDGMENTS}

The authors would like to thank the Brazilian agencies Capes, CNPq and FAPEPI
for partial financial support.\pagebreak%
\pagebreak%
%TCIMACRO{\FRAME{ftbpFU}{5.642in}{4.3664in}{0pt}{\Qcb{Phase diagram for the
%quantum transitions \ as a function of the long-range interaction $I/J_{1}$
%for $p=2$ and $J_{2}/J_{1}=0.2,2.0.$ For $J_{2}/J_{1}=0.2$ there are three
%phases, one spin liquid phase (QSL-I) and two saturated ferromagnetic
%phases(SF), and for $J_{2}/J_{1}=2.0,$ there are four phases, two spin liquid
%phases (QSL-I,QSL-II) and two saturated ferromagnetic phases(SF). The critical
%lines correspond to first-order phase transitions for $I/J_{1}>0$ and to
%second-order phase transitions for $I/J_{1}\leq0.$}}{\Qlb{fig01}}%
%{fig01.eps}{\special{ language "Scientific Word";  type "GRAPHIC";
%width 5.642in;  height 4.3664in;  depth 0pt;  original-width 11.4588in;
%original-height 8.8539in;  cropleft "0";  croptop "1";  cropright "1";
%cropbottom "0";  filename 'fig01.gif';file-properties "XNPEU";}}}%
%BeginExpansion
\pagebreak%
%EndExpansion
\pagebreak%

%TCIMACRO{\FRAME{ftbpFU}{5.642in}{4.3664in}{0pt}{\Qcb{Phase diagram for the
%quantum transitions \ as a function of the long-range interaction $I/J_{1}$
%for $p=2$ and $J_{2}/J_{1}=-0.2,-2.0.$ For $J_{2}/J_{1}=-0.2$ there are three
%phases, one spin liquid phase (QSL-I) and two saturated ferromagnetic
%phases(SF), and for $J_{2}/J_{1}=-2.0,$ there are four phases, two spin liquid
%phases (QSL-I,QSL-II) and two saturated ferromagnetic phases(SF). The critical
%lines correspond to first-order phase transitions for $I/J_{1}>0$ and to
%second-order phase transitions for $I/J_{1}\leq0.$}}{\Qlb{fig02}}%
%{fig02.gif}{\special{ language "Scientific Word";  type "GRAPHIC";
%width 5.642in;  height 4.3664in;  depth 0pt;  original-width 11.4588in;
%original-height 8.8539in;  cropleft "0";  croptop "1";  cropright "1";
%cropbottom "0";  filename 'fig02.gif';file-properties "XNPEU";}}}%
%BeginExpansion
%EndExpansion
\pagebreak%

%TCIMACRO{\FRAME{ftbpFU}{5.642in}{4.3664in}{0pt}{\Qcb{Magnetization $M^{z}$ as
%a function of $h/J_{1},$ for $p=2$, $J_{2}/J_{1}=2.0$ and different values
%$I/J_{1}.$ }}{\Qlb{fig03}}{fig03.gif}{\special{ language "Scientific Word";
%type "GRAPHIC";  maintain-aspect-ratio TRUE;  display "USEDEF";
%valid_file "F";  width 5.642in;  height 4.3664in;  depth 0pt;
%original-width 11.4588in;  original-height 8.8539in;  cropleft "0";
%filename 'fig03.gif';file-properties "XNPEU";}}}%
%BeginExpansion
%EndExpansion
\pagebreak%

%TCIMACRO{\FRAME{ftbpFU}{5.642in}{4.3664in}{0pt}{\Qcb{Isothermal susceptibility
%$\chi_{T}^{zz}$ as a function of $h/J_{1},$ for $p=2$, $J_{2}/J_{1}=2.0$ and
%different values $I/J_{1}.$}}{\Qlb{fig04}}{fig04.gif}%
%width 5.642in;  height 4.3664in;  depth 0pt;  original-width 11.4588in;
%original-height 8.8539in;  cropleft "0";  croptop "1";  cropright "1";
%cropbottom "0";  filename 'fig04.gif';file-properties "XNPEU";}}}%
%BeginExpansion
%EndExpansion
\pagebreak%

%TCIMACRO{\FRAME{ftbpFU}{5.0652in}{3.9219in}{0pt}{\Qcb{Global phase diagram for
%$p=2$ and $J_{2}/J_{1}\geq0$, as a function of $h/J_{1}$ and $I/J_{1}$.}%
%}{\Qlb{fig05}}{fig05.eps}{\special{ language "Scientific Word";
%type "GRAPHIC";  maintain-aspect-ratio TRUE;  display "USEDEF";
%valid_file "F";  width 5.0652in;  height 3.9219in;  depth 0pt;
%original-width 16.2498in;  original-height 12.5623in;  cropleft "0";
%filename 'fig05.eps';file-properties "XNPEU";}}}%
%BeginExpansion
%EndExpansion
\pagebreak%

%TCIMACRO{\FRAME{ftbpFU}{5.3558in}{4.1451in}{0pt}{\Qcb{Global phase diagram for
%$p=2$ and $J_{2}/J_{1}\leq0$, as a function of $h/J_{1}$ and $I/J_{1}$.}%
%}{\Qlb{fig06}}{fig06.eps}{\special{ language "Scientific Word";
%type "GRAPHIC";  maintain-aspect-ratio TRUE;  display "USEDEF";
%valid_file "F";  width 5.3558in;  height 4.1451in;  depth 0pt;
%original-width 17.1873in;  original-height 13.2809in;  cropleft "0";
%filename 'fig06.eps';file-properties "XNPEU";}}}%
%BeginExpansion
%EndExpansion
\pagebreak%

%TCIMACRO{\FRAME{ftbpFU}{5.642in}{4.3664in}{0pt}{\Qcb{ Phase diagram for the
%quantum transitions \ as a function of the long-range interaction $I/J_{1}$
%for $p\rightarrow\infty$ and $J_{2}/J_{1}=27/23,$ $2.0.$ For $J_{2}%
%/J_{1}=27/23$ there are three phases, one spin liquid phase (QSL-I) and two
%saturated ferromagnetic phases(SF), and for $J_{2}/J_{1}=2.0,$ there are four
%phases, two spin liquid phases (QSL-I,QSL-II) and two saturated ferromagnetic
%phases(SF). The critical lines correspond to first-order phase transitions for
%$I/J_{1}>0$ and to second-order phase transitions for $I/J_{1}\leq0.$ }%
%}{\Qlb{fig07}}{fig07.eps}{\special{ language "Scientific Word";
%type "GRAPHIC";  maintain-aspect-ratio TRUE;  display "USEDEF";
%valid_file "F";  width 5.642in;  height 4.3664in;  depth 0pt;
%original-width 11.4588in;  original-height 8.8539in;  cropleft "0";
%filename 'fig07.eps';file-properties "XNPEU";}}}%
%BeginExpansion
%EndExpansion
\pagebreak%

%TCIMACRO{\FRAME{ftbpFU}{5.642in}{4.3664in}{0pt}{\Qcb{ Phase diagram for the
%quantum transitions \ as a function of the long-range interaction $I/J_{1}$
%for $p\rightarrow\infty$ and $J_{2}/J_{1}=-0.2,-1/11.$ For $J_{2}/J_{1}=-0.2$
%there are four phases, two spin liquid phases (QSL-I,QSL-II) and two saturated
%ferromagnetic phases(SF), and for $J_{2}/J_{1}=-1/11,$ there are three phases,
%one spin liquid phase (QSL) and two saturated ferromagnetic phases(SF). The
%critical lines correspond to first-order phase transitions for $I/J_{1}>0$ and
%to second-order phase transitions for $I/J_{1}\leq0.$ }}{\Qlb{fig08}%
%}{fig08.eps}{\special{ language "Scientific Word";  type "GRAPHIC";
%width 5.642in;  height 4.3664in;  depth 0pt;  original-width 11.4588in;
%original-height 8.8539in;  cropleft "0";  croptop "1";  cropright "1";
%cropbottom "0";  filename 'fig08.eps';file-properties "XNPEU";}}}%
%BeginExpansion
%EndExpansion
\pagebreak%

%TCIMACRO{\FRAME{ftbpFU}{5.642in}{4.3664in}{0pt}{\Qcb{ Phase diagram for the
%quantum transitions \ as a function of the long-range interaction $I/J_{1}$
%for $p\rightarrow\infty$ and $J_{2}/J_{1}=-2.5,-1.7370...$ For $J_{2}%
%/J_{1}=-2.5$ there are four phases, two spin liquid phases (QSL-I,QSL-II) and
%two saturated ferromagnetic phases(SF), and for $J_{2}/J_{1}=-1.7370..,$ there
%are four phases, two spin liquid phases (QSL-I,QSL-II)) and two saturated
%ferromagnetic phases(SF) four phases which coexist in the quadruple point
%localized at $h/J_{1}=0,$ $I/J_{1}=1.7798...$ The functional of the free
%energy for this case is shown in Fig. \ref{fig11}. The critical lines
%correspond to first-order phase transitions for $I/J_{1}>0$ and to
%second-order phase transitions for $I/J_{1}\leq0.$ }}{\Qlb{fig09}}%
%{fig09.eps}{\special{ language "Scientific Word";  type "GRAPHIC";
%width 5.642in;  height 4.3664in;  depth 0pt;  original-width 11.4588in;
%original-height 8.8539in;  cropleft "0";  croptop "1";  cropright "1";
%cropbottom "0";  filename 'fig09.eps';file-properties "XNPEU";}}}%
%BeginExpansion
%EndExpansion
\pagebreak%

%TCIMACRO{\FRAME{ftbpFU}{5.642in}{4.3664in}{0pt}{\Qcb{ Phase diagram for the
%quantum transitions \ as a function of the long-range interaction $I/J_{1}$
%for $p\rightarrow\infty$ and $J_{2}/J_{1}=-3.5,-3.0.$ For $J_{2}/J_{1}=-3.5$
%there are four phases, two spin liquid phases (QSL-I,QSL-II) and two saturated
%ferromagnetic phases(SF), and for $J_{2}/J_{1}=-3.0,$ there are three phases,
%one spin liquid phase (QSL-I) and two saturated ferromagnetic phases(SF).The
%critical lines correspond to first-order phase transitions for $I/J_{1}>0$ and
%to second-order phase transitions for $I/J_{1}\leq0.$ }}{\Qlb{fig10}%
%}{fig10.eps}{\special{ language "Scientific Word";  type "GRAPHIC";
%width 5.642in;  height 4.3664in;  depth 0pt;  original-width 11.4588in;
%original-height 8.8539in;  cropleft "0";  croptop "1";  cropright "1";
%cropbottom "0";  filename 'fig10.eps';file-properties "XNPEU";}}}%
%BeginExpansion
%EndExpansion
\pagebreak%

%TCIMACRO{\FRAME{ftbpFU}{5.642in}{4.3664in}{0pt}{\Qcb{Functional of the free
%energy $f$ as a function of the magnetization $M^{z}$ for $p\rightarrow\infty
%$, $h/J_{1}=0,$ $I/J_{1}=1.7798...$ and $J_{2}/J_{1}=-1.7370...$ at the
%quadruple point.}}{\Qlb{fig11}}{fig11.eps}%
%width 5.642in;  height 4.3664in;  depth 0pt;  original-width 11.4588in;
%original-height 8.8539in;  cropleft "0";  croptop "1";  cropright "1";
%cropbottom "0";  filename 'fig11.eps';file-properties "XNPEU";}}}%
%BeginExpansion
%EndExpansion
\pagebreak%

%TCIMACRO{\FRAME{ftbpFU}{5.642in}{4.3664in}{0pt}{\Qcb{ Phase diagram for the
%quantum transitions as a function of $J_{2}/J_{1}$ for $p=3$ and $I/J_{1}=0$.
%All the transitions are of second-order and there are five phases, three spin
%liquid phase (QSL-I, QSL-II, QSL-III) and two saturated ferromagnetic phases
%(SF).}}{\Qlb{fig12}}{fig12.eps}{\special{ language "Scientific Word";
%type "GRAPHIC";  maintain-aspect-ratio TRUE;  display "USEDEF";
%valid_file "F";  width 5.642in;  height 4.3664in;  depth 0pt;
%original-width 11.4588in;  original-height 8.8539in;  cropleft "0";
%filename 'fig12.eps';file-properties "XNPEU";}}}%
%BeginExpansion
%EndExpansion
\pagebreak%
%TCIMACRO{\FRAME{ftbpFU}{5.9343in}{4.2886in}{0pt}{\Qcb{Phase diagram for the
%quantum transitions as a function of $J_{2}/J_{1}$ for $p=4$ and $I/J_{1}=0$.
%All the transitions are of second-order and there are five phases, three spin
%liquid phase (QSL-I, QSL-II, QSL-IV) and two saturated ferromagnetic phases
%(SF).}}{\Qlb{fig13}}{fig13.eps}{\special{ language "Scientific Word";
%type "GRAPHIC";  maintain-aspect-ratio TRUE;  display "USEDEF";
%valid_file "F";  width 5.9343in;  height 4.2886in;  depth 0pt;
%original-width 9.8329in;  original-height 7.0941in;  cropleft "0";
%filename 'fig13.eps';file-properties "XNPEU";}}}%
%BeginExpansion
%EndExpansion
\pagebreak%

%TCIMACRO{\FRAME{ftbpFU}{5.54in}{4.2869in}{0pt}{\Qcb{Static correlation
%function $\langle S_{j}^{x}S_{j+r}^{x}\rangle$as a function of $r$, for $p=2$,
%$h/J_{1}=2.0,$ $J_{2}/J_{1}=2.0$ and $I/J_{1}=-1.0.$ The inset shows the power
%law decay of the correlation characteristic of the QSL-I phase.}}{\Qlb{fig14}%
%}{fig14.eps}{\special{ language "Scientific Word";  type "GRAPHIC";
%width 5.54in;  height 4.2869in;  depth 0pt;  original-width 11.4588in;
%original-height 8.8539in;  cropleft "0";  croptop "1";  cropright "1";
%cropbottom "0";  filename 'fig14.eps';file-properties "XNPEU";}}}%
%BeginExpansion
%EndExpansion
\pagebreak%
%TCIMACRO{\FRAME{ftbpFU}{5.54in}{4.2869in}{0pt}{\Qcb{Static correlation
%function $\langle S_{j}^{x}S_{j+r}^{x}\rangle$as a function of $r$, for $p=2$,
%$h/J_{1}=-1.0,$ $J_{2}/J_{1}=2.0$ and $I/J_{1}=-1.0.$ The inset shows the
%power law decay of the correlation characteristic of QSL-II phase with the
%oscillatory modulation\QTR{bf}{ }$f(r).$\QTR{bf}{ }}}{\Qlb{fig15}}%
%{fig15.eps}{\special{ language "Scientific Word";  type "GRAPHIC";
%width 5.54in;  height 4.2869in;  depth 0pt;  original-width 11.4588in;
%original-height 8.8539in;  cropleft "0";  croptop "1";  cropright "1";
%cropbottom "0";  filename 'fig15.eps';file-properties "XNPEU";}}}%
%BeginExpansion
%EndExpansion
\pagebreak%

%TCIMACRO{\FRAME{ftbpFU}{5.54in}{4.2869in}{0pt}{\Qcb{Static correlation
%function $\langle S_{j}^{x}S_{j+r}^{x}\rangle$as a function of $r$, for $p=3$,
%$h/J_{1}=-0.15,$ $J_{2}/J_{1}=3.0$ and $I/J_{1}=0.$ The inset shows the power
%law decay of the correlation characteristic of QSL-III phase with the
%oscillatory modulation \QTR{bf}{ }$f(r).$}}{\Qlb{fig16}}{fig16.eps}%
%width 5.54in;  height 4.2869in;  depth 0pt;  original-width 11.4588in;
%original-height 8.8539in;  cropleft "0";  croptop "1";  cropright "1";
%cropbottom "0";  filename 'fig16.eps';file-properties "XNPEU";}}}%
%BeginExpansion
%EndExpansion
\pagebreak%

%TCIMACRO{\FRAME{ftbpFU}{5.54in}{4.2869in}{0pt}{\Qcb{Static correlation
%function $\langle S_{j}^{x}S_{j+r}^{x}\rangle$as a function of $r$, for $p=4$,
%$h/J_{1}=-0.6,$ $J_{2}/J_{1}=1.6$ and $I/J_{1}=0.$ The inset shows the power
%law decay of the correlation characteristic of QSL-IV phase with the
%oscillatory modulation\QTR{bf}{ }$f(r).$\QTR{bf}{ }}}{\Qlb{fig17}}%
%{fig17.eps}{\special{ language "Scientific Word";  type "GRAPHIC";
%width 5.54in;  height 4.2869in;  depth 0pt;  original-width 11.4588in;
%original-height 8.8539in;  cropleft "0";  croptop "1";  cropright "1";
%cropbottom "0";  filename 'fig17.eps';file-properties "XNPEU";}}}%
%BeginExpansion
%EndExpansion


\begin{thebibliography}{99}                                                                                               %


\bibitem {Schadev 2000}S. Sachdev, \textit{Quantum Phase Transitions}
(Cambrigde University Press, Cambrigde, England, 2000).

\bibitem {Coleman 2005}P. Coleman and A. Schofield, Nature \textbf{20,
}vol-433 (2005).

\bibitem {Schadev 2008}S. Schadev, Nature Physics, vol-\textbf{4} (2008)

\bibitem {Raoul 2008}R. Dillenschneider, Phys. Rev. B \textbf{78, }224413 (2008).

\bibitem {Carollo 2005}A. C. M. Carollo and J. K. Pachos, Phys. Rev. Lett.
\textbf{95,} 157203 (2005).

\bibitem {Zhu 2006}S. L. Zhu, Phys. Rev. Lett. \textbf{96}, 077206 (2006).

\bibitem {Zanardi 2006}P. Zanardi and N. Paunkovic, Phys. Rev. E \textbf{74},
031123 (2006).

\bibitem {Venuti 2007}L. C. Venuti and P. Zanardi, Phys. Rev. Lett.
\textbf{99}, 095701 (2007).

\bibitem {Amico 2006}L. Amico, F. Baroni, A. Fubini, D. Patan\`{e}, V.
Tognetti, and P. Verrucchi, Phys. Rev. A \textbf{74}, 022322 (2006).

\bibitem {Lieb 1961}H. E. Lieb, T. Schultz, and D. C. Mattis, Ann. Phys. (N.
Y.) \textbf{16}, 407 (1961).

\bibitem {Titvinidze 2003}I. Titvinidze and G. I. Japaridze, Eur. Phys. J. B,
\textbf{32}, 383-393 (2003).

\bibitem {Krokhmalskii2008}T. Krokhmalskii, O. Derzhko, J. Stolze, and T.
Verkholyak, Phys. Rev. B,\textbf{ 77}, 174404 (2008).

\bibitem {Lenilson 2005}L. L. Gon\c{c}alves , L. P. S. Coutinho and J. P. de
Lima , Physica A\textit{,\textbf{ }}\textbf{345}\textit{, }71-91\textit{ }(2005).

\bibitem {Bouchett 2008}F. Bouchett, T. Dauxois, D. Mukamel, and S. Ruffo,
Phys. Rev. E \textbf{77}, 011125 (2008).

\bibitem {Derzhko 2009}O. Derzhko, T. Krokhmalskii, J. Stolze, and T.
Verkholyak, Phys. Rev. B \textbf{79}, 094410 (2009).

\bibitem {de Lima 2008}J. P. de Lima and L.L. Gon\c{c}alves, Phys. Rev. B
\textbf{77}, 214424 (2008).

\bibitem {Pfleiderer 2005}C. Pfleiderer, J. Phys: Condens. Matter \textbf{17},
S987 (2005)

\bibitem {Suzuki 1971}M. Suzuki, Phys. Lett. A \textbf{34}, 94 (1971); Prog.
Theor. Phys. \textbf{46}, 1337 (1971)

\bibitem {Continentino 2004}M. A. Continentino and A. S. Ferreira, Physica A
\textbf{339}, 461 (2004).

\bibitem {Lee 2008}P. Lee, Science\textit{, }\textbf{321}, 1306-1307 (2008).

\bibitem {siskens 1974}Th. J. Siskens and P. Mazur, \textit{Physica}
\textbf{71, }560 (1974).

\bibitem {capel 1977}H. W. Capel and J. H. H. Perk, \textit{Physica A}
\textbf{87, }211 (1977).

\bibitem {goncalves 1977}L. L. Gon\c{c}alves, \textit{Theory of properties of
some one-dimensional systems} (D.\ Phil. Thesis, University of Oxford, 1977)

\bibitem {amit1984}D. Amit, \textit{Field Theory}, \textit{The Renormalization
Group, and Critical Phenomena}, World Scientific, Singapore, 1984.

\bibitem {murray 1984}J. D.Murray, \textit{Asymptotic Analysis, Applied
Mathematical Sciences}, Vol 48 (Springer-Verlag, Berlim 1984).

\bibitem {Mattuck 1992}R. D. Mattuck, \textit{A Guide to Feynman Diagrams in
the Many-Body Problem} (Dover Publications, New York, 1992).

\bibitem {McCoy 1971}E. Barouch and B. M. McCoy, Phys. Rev. A 3, 786 (1971).

\bibitem {Stanley 1971}H.E. Stanley, \textit{Phase Transitions and Critical
Phenomena} (Oxford University Press, Oxford, 1971)$.$

\bibitem {Lima 1994}J. P. de Lima and L. L. Gon\c{c}alves, Mod. Phys. Lett. B
8, 871 (1994).

\bibitem {Continentino 2001}M. A. Continentino, \textit{Quantum Scaling in
Many-Body Systems }(World Scientific Publishing, 2001).
\end{thebibliography}
\end{document}